\RequirePackage{amsmath}
\documentclass[11pt,onecolumn,a4paper]{article}
\usepackage[utf8]{inputenc}
\usepackage[english]{babel}
\usepackage[latexsym]{}
\usepackage{amsmath}
\usepackage{amsfonts}

\usepackage{geometry}
\usepackage{graphicx}
\usepackage{float}
\usepackage{amsmath}
\usepackage{amssymb,amsthm}
\usepackage{wasysym}
\usepackage[square,sort,comma,numbers]{natbib}

\theoremstyle{plain}
\newtheorem{thm1}{Theorem}[section]
\newtheorem{thm2}{Theorem}[section]

\theoremstyle{definition}
\newtheorem{definition}[thm1]{Definition} 
\newtheorem{example}[thm2]{Example} 

\title{RDF 1.1: Knowledge Representation and Data Integration Language for the Web}

\author{Dominik Tomaszuk, and David Hyland-Wood}

\date{}

\begin{document}

\maketitle

\framebox[1.1\width]{Submitted to Symmetry https://doi.org/10.3390/sym12010084} \par

\abstract{Resource Description Framework (RDF) can seen as a solution in today's landscape of knowledge representation research. An RDF language has symmetrical features because subjects and objects in triples can be interchangeably used. Moreover, the regularity and symmetry of the RDF language allow knowledge representation that is easily processed by machines, and because its structure is similar to natural languages, it is reasonably readable for people. RDF provides some useful features for generalized knowledge representation. Its distributed nature, due to its identifier grounding in IRIs, naturally scales to the size of the Web. However, its use is often hidden from view and is, therefore, one of the less well-known of the knowledge representation frameworks. Therefore, we summarise RDF v1.0 and v1.1 to broaden its audience within the knowledge representation community. This article reviews current approaches, tools, and applications for mapping from relational databases to RDF and from XML to RDF. We discuss RDF serializations, including formats with support for multiple graphs and we analyze RDF compression proposals. Finally, we present a summarized formal definition of RDF 1.1 that provides additional insights into the modeling of reification, blank nodes, and entailments.}

\section{Introduction}
\label{sec:introduction}

The Resource Description Framework version 1.1 is a modern and complete knowledge representation framework that is seemingly underrepresented within the traditional knowledge representation research community. We seek to clarify some differences between the way RDF 1.1 was defined in World Wide Web Consortium (W3C) specifications and the ways in which it is reinterpreted during implementation. Firstly, we need to discuss how RDF relates to the broader field of knowledge representation.

Knowledge representation can be seen as the way in which knowledge is presented in a language. More precisely it was clarified by 
Sowa \cite{Sowa1999}, who  presents five characteristics of knowledge representation:
\begin{enumerate}
 \item It is most fundamentally a surrogate.
 \item It is a collection of ontological commitments.
 \item It is a fragmental intelligent reasoning theory.
 \item It is a pragmatically efficient computation medium.
 \item It is a human expression medium.
\end{enumerate}

Natural language can be defined as one of the methods of knowledge representation. The fundamental unit of knowledge in 
such languages is often a sentence that consists of a set of words arranged according to grammatical rules. In spite of the 
existence of grammatical rules that encode expectations of word order, irregularities and exceptions to the rules make it 
difficult for machines to process natural languages.

The RDF data model was a response to this problem for knowledge representation on the World Wide Web. This language and the 
notions from which it originates have enabled free data exchange, formalization, and unification of stored knowledge. RDF 
was developed iteratively over nearly two decades to address knowledge representation problems at Apple Computer, Netscape 
Communications Corporation, and the Semantic Web and Linked Data projects at the World Wide Web Consortium.

An assumption in RDF \cite{Raimond2014} is to define resources by means of the statement consisting of three elements (the so-called RDF \emph{triple}): \emph{subject}, \emph{predicate}, and \emph{object}.
RDF borrows strongly from natural languages. An RDF \emph{triple} 
may then be seen as an expression with \emph{subject} corresponding to the subject of a sentence, \emph{predicate} 
corresponding to its verb and \emph{object} corresponding to its object \cite{schwitter2004}. So the RDF language may be categorized into the same syntactic criteria as natural languages. According to these premises, RDF belongs to the group of Subject 
Verb Object languages (SVO) \cite{Crystal1997}. The consistency and symmetry of the RDF language allows knowledge representation 
that is easily processed by machines, and because its structure is similar to that of natural languages, it is 
reasonably readable for people.

On the other hand, following Lenzerini \cite{lenzerini2002}, data integration is the issue of joining data 
stored at disparate sources, and providing the user with an integrated perspective of these data. Much of the data on 
the Web is stored in relational databases. A similar amount of data exists in hierarchical files such as XML documents. 
Integration of all of this data would provide huge profits to the organizations, enterprises, and governments that own 
the data.

Interoperability is the capability of two or more (different) software systems or their components to interchange information and to use the information that has been shared \cite{ceri1991}. In the context of the Web, interoperability is concerned with the support of applications that exchange and share information across the boundaries of existing data sources. The RDF world is a satisfying method for organizing information from different information sources. In particular, RDF can be seen as a general proposition language for the Web, which consolidates data from heterogeneous sources. It can provide interoperability between applications that exchange the data.

Knowledge representation and data integration in the context of RDF is relevant for several reasons, including: promotes data exchange and interoperability; facilitates the reuse of available systems and tools; enables a fair comparison of Web systems by using benchmarks. In particular, this article shows how the RDF data model can be related to other models.

The RDF language enables large portions of existing data to be processed and analyzed. This produces the need to develop the foundations of this language. This article addresses this challenge by developing an abstract model that is suitable to formalize and explain properties about the RDF data. We study the RDF data model, minimal and maximal representations, and show complexity bounds for the main problems.

\subsection{Contributions}
When we examine the state of the RDF data model, we see evidence of trade-offs that occurred as various constituencies 
took part in the design process. Many of these trade-offs were never completely summarized in the RDF standards. Our article reviews a final state of RDF, and to identify areas where this data model is poorly understood. Our contributions are:
\begin{enumerate}
 \item to compare the RDF reification approaches,
 \item to analyze the RDF~1.1 interpretations, entailments and their complexity,
 \item to study the RDF blank nodes and their complexity,
 \item to compare the various RDF data integration approaches,
 \item to compare the RDF~1.1 serialization formats, including multiple graph syntaxes,
 \item to compare the RDF binary and compression formats.
\end{enumerate}
In addition, one of our main aims is to isolate a core fragment of RDF which is the point of a formal analysis. We
discuss a formalization of an RDF syntax and semantics, which is beneficial, for example, to help identify relationships 
among the constructors, to assist in optimization and validations of software implementations, and to identify redundant notions.

\subsection{Review Organization}
The remainder of this article is as follows: Section~\ref{sec:preliminaries} presents related work and formalized concepts for RDF. In Section~\ref{sec:bnodes} we discuss RDF blank
nodes and their complexity. Section~\ref{sec:entailments} analyzes a semantics for the RDF and, outlines a set of different 
entailment regimes. Section~\ref{sec:integration} overviews and compares various proposals for RDF data integration. 
Section~\ref{sec:serializations} briefly introduces and compares various serialization formats for RDF~1.1. 
In Section~\ref{sec:compression} we overview and compare various proposals for RDF compression. Finally, 
Section~\ref{sec:conclusions} gives some concluding remarks.

\section{Literature Review}
\label{sec:preliminaries}

In this Section, we present chronologically related works, and show a formalized syntax and concept for RDF.

The pre-version of RDF was published in 1999 \cite{Lassila1999}. In this document RDF did not have many features known from current versions, e.g. there were no explicit blank nodes. 
One of the first paper is published in 2000 \cite{decker2000}, which present RDF and RDFS. In 2001 Champin \cite{champin2001} 
focuses on RDF model and XML syntax of RDF. In \cite{carroll2002} Carroll provides a formal analysis of comparing RDF graphs. 
An author proves that isomorphism of an RDF graph can be reduced to known graph isomorphism problems. In \cite{Pan2003}, the 
authors focus on delineating RDFS(FA) semantics for RDF Schema, which can interoperate with common first-order languages. 
Grau \cite{grau2004} continues RDF(FA) approach and proposes a possible simplification of the Semantic Web 
architecture. Yang \textit{et al.} \cite{yang2003} propose a semantics for anonymous resources and statements in 
F-logic \cite{kifer1989}. Another overview is presented in \cite{berners2003} by Berners-Lee and in \cite{Marin2004} by Marin. 

The RDF~1.0 recommendation \cite{Manola2004} has been reviewed of several analyzes. In 2004 
Gutierrez \textit{et al.} \cite{gutierrez2004} formalized RDF and investigate computational aspects of testing entailment and 
redundancy. In 2005, in \cite{bruijn2005,franconi2005}, the authors propose a logical reconstruction of the RDF family 
languages. Feigenbaum \cite{feigenbaum2007} briefly describes RDF with emphasis on the Semantic Web. 
In \cite{Munoz2007,munoz2009}, the authors discuss fragments of RDF and systematize them. These papers also outline 
complexity bound for ground entailment in the proposed fragment. Yet another approach is provided in \cite{Pichler2008}, 
which presents domain-restricted RDF (dRDF). In 2011 and 2012 another descriptions are shown in \cite{Hitzler2011} and 
\cite{antoniou2012}, which presents Semantic Web technologies. In 2014 Curé \textit{et al.} \cite{cure2014} briefly introduce RDF 1.0 and RDFS 1.0.  There are also a lot of papers that extend to RDF 
annotations \cite{straccia2010,buneman2010,udrea2006} e.g. for fuzzy metrics \cite{straccia2009}, temporal 
metrics \cite{gutierrez2005}, spatial metrics \cite{koubarakis2010} and trust metrics \cite{tomaszuk2012}. A separate group are publications on data integration in the context of RDF \cite{sequeda2009,spanos2012,harth2014,thakkar2019}. The term relational database to RDF mapping has been used in \cite{spanos2012}. In \cite{sequeda2009,thakkar2019} the authors propose direct mappings, and in \cite{harth2014} indirect mappings is presented.

In order for machines to exchange machine-readable data, they need to agree upon a universal data model under which to 
structure, represent and store content. This data model should be general enough to provide representation 
for arbitrary data content regardless of its structure. The data model should also enable the processing of this content. The core data 
model selected, for use on the Semantic Web and Web of Data digital ecosystems is RDF.

RDF constitutes a common method of the conceptual description or information modeling accessible in Web resources.
It provides the crucial foundation and framework to support the description and management of data. 
Especially, RDF is a general data model for resources and relationship descriptions between them.

The RDF data model rests on the concept of creating web-resource statements in the form of subject-predicate-object expressions, 
which in the RDF terminology, are referred to as \emph{triples}.

An RDF triple consists of a subject, a predicate, and an object. In \cite{Cyganiak2014}, the meaning of subject,
predicate and object is clarified. The \emph{subject} expresses a resource, the \emph{object} fills the value of the relation, 
the \emph{predicate} refers to the features or aspects of resource and expresses a subject-object relationship. 
The predicate indicates a binary relation, also known as a property.

Following \cite{Cyganiak2014}, we provide definitions of RDF triples below.

\begin{definition}[RDF triple]
\label{def:rdftriple}
Assume that $\mathcal{I}$ is the set of all Internationalized Resource Identifiers (IRIs), 
$\mathcal{B}$ (an infinite) set of blank nodes, $\mathcal{L}$ the set of literals.
An \emph{RDF triple} is defined as a triple $t = \langle s, p, o\rangle$ where $s \in \mathcal{I} \cup \mathcal{B}$ is 
called the \emph{subject}, $p \in \mathcal{I}$ is called the \emph{predicate} and $o \in \mathcal{I} \cup \mathcal{B} \cup \mathcal{L}$ 
is called the \emph{object}.\qed
\end{definition}

\begin{example}
The example presents an RDF triple consisting of subject, predicate and object.
\begin{verbatim}
<http://example.com/me#john>   foaf:firstName   "John" .
\end{verbatim}
\label{ex:rdftripleex}
\end{example}

The primitive constituents of the RDF data model are terms that can be utilized in reference to resources: 
anything with identity. The set of the terms is divided into three disjoint subsets: 
\begin{itemize}
 \item IRIs,
 \item blank nodes,
 \item literals.
\end{itemize}

\begin{definition}[IRIs]
\label{def:rdfterms}
\emph{IRIs} are a set of Unicode names in registered namespaces and addresses referring to registered protocols 
or namespaces used to identify a web resource. For example,
\texttt{<http://dbpe\\dia.org/resource/House>} is used to identify the house in DBpedia \cite{Auer2007}.\qed
\end{definition}
Note that in RDF 1.0 identifiers were RDF URI (Uniform Resource Identifier) References. Identifiers in RDF~1.1 are now IRIs, which are URIs
generalization, which allows a wider range of Unicode codes. Please notice that each absolute URL (and URI) is 
an Internationalized Resource Identifier, but not every IRI is a URI. When one is implemented in operations 
that are exclusively specified for URI, it should first be transformed.

IRIs can be shortened. RDF syntaxes use two similar ways: CURIE (compact URI expressions) \cite{Birbeck2010} 
and QNames (qualified names) \cite{Bray2006}. Both are comprised of two components: an optional prefix and 
a reference. The prefix is separated from the reference by a colon. The syntax of QNames is restrictive and 
does not allow all possible IRIs to be represented, i.e. \texttt{issn:15700844} is valid CURIE but invalid
QName. Syntactically, QNames are a subset of CURIEs.

\begin{definition}[Literals]
\label{def:rdfliterals}
\emph{Literals} are a set of lexical forms and datatype IRIs. A lexical form is a Unicode string, and a datatype IRI identifies an attribute of data that defines how the user intends to use the data. RDF borrows many of the datatypes defined in XML Schema~1.1 \cite{Sperberg-McQueen2013}.
For example, \texttt{"1"\textasciicircum\textasciicircum http://www.w3.org/2001/XMLSchema{\#}unsignedInt}, where \texttt{1} is a lexical
form and should be treated as unsigned integer number.\qed
\end{definition}
Note that in RDF~1.0 literals with a language tag do not support a datatype URI. In RDF~1.1 literals with language tags 
have the datatype IRI \texttt{rdf:langString}. In current version of RDF all literals have datatypes. Implementations might choose to 
support syntax for literals that have lexical form only, but it should be treated as synonyms for 
\texttt{xsd:string} literals. Moreover, RDF~1.1 may support the new datatype \texttt{rdf:HTML}. Both \texttt{rdf:HTML} 
and \texttt{rdf:XMLLiteral} depend on DOM4 (Document Object Model level 4)\footnote{DOM4 is an interface that treats an XML or HTML elements as objects, see http://www.w3.org/TR/dom/}. 

\begin{definition}[Blank nodes]
\label{def:bnodes}
\emph{Blank nodes} are defined as elements of an infinite set disjoint from IRIs and Literals.\qed
\end{definition}
In RDF~1.1 blank node identifiers are local identifiers adapted in particular RDF serializations or implementations of an RDF store.

A set of RDF triples represents a labeled directed multigraph. The nodes are the subjects and objects of 
their triples. RDF is also related to as being \emph{graph structured data} where each $\langle s, p, o\rangle$ triple can be 
interpreted as an edge $s \xrightarrow{p} o$.

\begin{definition}[RDF graph]
\label{def:rdfgraph}
Let $\mathcal{O} = \mathcal{I} \cup \mathcal{B} \cup \mathcal{L}$ and
$\mathcal{S} = \mathcal{I} \cup \mathcal{B}$. $G \subset \mathcal{S} \times \mathcal{I} \times \mathcal{O}$ 
is a finite subset of \emph{RDF triples} and called an \emph{RDF graph}.\qed
\end{definition}

\begin{figure}[ht]
\centering
\includegraphics{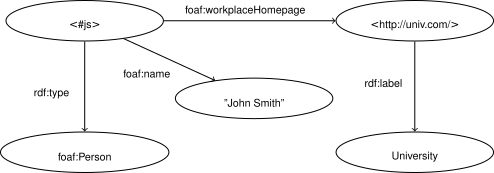}
\caption[An RDF graph]{An RDF graph with four triples.}
\label{fig:rdfgraph}
\end{figure}

\begin{example}
The example in Fig.~\ref{fig:rdfgraph} presents an RDF graph of a FOAF \cite{Brickley2014b} profile. This graph
includes four RDF triples:
\begin{verbatim}
<#js>   rdf:type    foaf:Person  .
<#js>   foaf:name   "John Smith" .
<#js>   foaf:workplaceHomepage <http://univ.com/> .
<http://univ.com/>  rdfs:label "University"       .
\end{verbatim}
\label{ex:rdfgraphex}
\end{example}

The RDF syntax and semantics can be widened to named graphs \cite{carroll2005}. The named graph data model is 
a variation of the RDF data model. The basic concept of the model consists of proposing a graph naming mechanism. 

\begin{figure}[ht]
\centering
\includegraphics{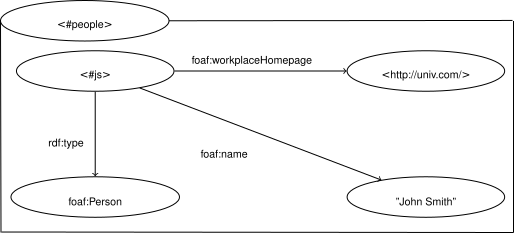}
\caption[A named graph]{A named graph identified by \texttt{<\#people>} with three triples.}
\label{fig:namedgraph}
\end{figure}

\begin{definition}[Named graph]
A \emph{named graph} $NG$ is a pair $\langle u, G \rangle$, where $u \in \mathcal{I} \cup \mathcal{B}$ is a graph name 
and $G$ is an RDF graph.\qed
\end{definition}
\begin{example}
The example in Fig.~\ref{fig:namedgraph} presents a named graph of a FOAF profile. This graph has the name 
\texttt{http://example.com/\#people} and includes three RDF triples:
\begin{verbatim}
<#people> {
 <#js>  rdf:type   foaf:Person  .
 <#js>  foaf:name  "John Smith" .
 <#js>  foaf:workplaceHomepage <http://univ.com/> .
}
\end{verbatim}
\label{ex:namedgraphex1}
\end{example}

RDF~1.1 describes the idea of RDF datasets, a collection of a distinguished RDF graph and zero or more graphs with context.Whereas RDF graphs have a formal semantics that establishes what arrangements of the universe make an RDF graph true, no agreed model-theoretic semantics exists for RDF datasets. For more about the above characteristics, we refer the interested reader to the \emph{RDF~1.1: On 
Semantics of RDF Datasets} \cite{Zimmermann2014} which specify several semantics in terms of model theory.

\begin{definition}[RDF dataset]
\label{def:rdfdataset}
An \emph{RDF dataset} $DS$ includes one nameless RDF graph, called the default graph and zero or more named graphs, where each is identified 
by IRI or blank node, $DS =  \{ G, \langle u_1, G_1 \rangle, \langle u_2, G_2 \rangle, \dots, \langle u_i, G_i \rangle \}$.\qed
\end{definition}
\begin{example}
The example presents an RDF dataset consisting of one unnamed (default) graph and two named graphs.
\begin{verbatim}
{ }  # empty RDF graph
ex:g1 { ex:s  ex:p  ex:o }
ex:g2 { ex:x  ex:y  ex:z }
\end{verbatim}
\label{ex:namedgraphex2}
\end{example}

In addition, the RDF Schema Recommendation \cite{Brickley2014a} provides a set of built-in vocabulary terms under a core RDF namespace that 
unifies popular RDF patterns, such as RDF collections, containers, and RDF reification.

RDF is a provider of vocabularies for container description. Each container has a type; what is more, their members can be 
itemized with the use of a fixed set of container membership properties. In order to provide a way to separate the members 
from one another, the properties must be indexed by integers, however, these indexes can not be regarded as specifying 
an ordering of the RDF container itself. The RDF containers are RDF graph entities that use the vocabulary in order 
to provide basic information about the entities and give a description of the container members. Following 
\cite{Brickley2014a}, RDF gives vocabularies for specifying three container classes:
\begin{itemize}
 \item \texttt{rdf:Bag} is unordered container and allows duplicates,
 \item \texttt{rdf:Seq} is ordered container,
 \item \texttt{rdf:Alt} is considered to define a group of alternatives.
\end{itemize}
\begin{example}
The example presents an \texttt{rdf:Bag} container representing the group of resources.
\begin{verbatim}
<http://example.com/p> ex:teachers _:a .
_:a rdf:type rdf:Bag .
_:a rdf:_1 <http://example.com/p/js> .
_:a rdf:_2 <http://example.com/p/ak> .
\end{verbatim}
\label{ex:container}
\end{example}

Another feature of RDF is its vocabulary for describing RDF collections. Since the RDF data model has no inherent ordering, collections can be used to determine an ordered and linear collection using a linked list pattern. RDF collections are in the form 
of a linked list structure such that it comprises of elements with a member and a pointer to the next element. Moreover, 
collections are closed lists in contrast to containers, which allow the set of items in the group to be precisely 
determined by applications. However cyclic or unterminated lists in RDF are possible.
\begin{example}
The example presents a collection representing the group of resources. In the graphs, each member of the collection is the object
of the \texttt{rdf:first} predicate whose subject is a blank node representing a list, that links by the \texttt{rdf:rest}
predicate. The \texttt{rdf:rest} predicate, with the \texttt{rdf:nil} resource as its object, indicates the end of the list.
\begin{verbatim}
<http://example.com/p> ex:teachers _:x .
_:x rdf:first <http://example.com/p/js> .
_:x rdf:rest _:y .
_:y rdf:first <http://example.com/p/ak> .
_:y rdf:rest rdf:nil .
\end{verbatim}
\label{ex:collection}
\end{example}

Another RDF feature is reification (denoted sr in Table~\ref{tab:reicomp}), which provides an approach to talk about individual RDF triples themselves 
within RDF. The method allows for constructing a new resource that refers to a triple, and then for adding supplementary 
information about that RDF statement.
\begin{example} 
The example presents a standard RDF reification.
\begin{verbatim}
ex:t rdf:type      rdf:Statement .
ex:t rdf:subject   <#js> .
ex:t rdf:predicate foaf:name . 
ex:t rdf:object    "John Smith" .
ex:t ex2:certainty 0.5 .
\end{verbatim}
\label{ex:reification}
\end{example}

An extension of the previous method is N-ary Relations~\cite{Noy2006} (denoted nr in Table~\ref{tab:reicomp}). This approach is not strictly designed for reification, but focuses on additional arguments in the relation to provide extra information about the relation instance itself.
\begin{example} 
The example presents an N-ary Relations.
\begin{verbatim}
<#js> foaf:name "John Smith" .
<#js> ex2:certainty _:r .
_:r ex2:certainty-value 0.5 .
\end{verbatim}
\label{ex:nrel}
\end{example}

There are other proposals  \cite{hartig2014,nguyen2014} of RDF reification. The first proposal called RDF*/RDR\footnote{Also known 
as Reification Done Right (RDR)} \cite{hartig2014} (denoted rdr in Table~\ref{tab:reicomp}) is an alternative approach to represent statement-level metadata. It is based 
on the idea of using a triple in the subject or object positions of other triples that represent metadata about the embedded statement. 
To reified RDF data an additional file format based on Turtle has been introduced. Unfortunately, in RDF*/RDR several reification 
statements of the same triple are translated into one standard reification part so that, it is not possible to distinguish 
grouped annotations.
\begin{example}
The example presents an RDF*/RDR.
\begin{verbatim}
<<<#js> foaf:name "John Smith">> ex2:certainty 0.5 .
\end{verbatim}
\label{ex:rdfstar}
\end{example}
The second proposal is called Singleton Property \cite{nguyen2014} (denoted sp in Table~\ref{tab:reicomp}). It is for representing statements about statements. 
It uses a unique predicate for every triple with associated metadata to the statement, which can be linked to the high-level predicate. 
Authors propose special predicate \texttt{singletonPropertyOf} to link to original predicate. Since the predicate resource use 
predicate \texttt{singletonPropertyOf}, it is possible to use RDFS entailment rules to infer the original statements.
\begin{example}
The example presents a Singleton Property.
\begin{verbatim}
foaf:name#1 rdf:singletonPropertyOf foaf:name .
<#js> foaf:name#1 "John Smith" .
foaf:name#1 ex2:certainty 0.5 .
\end{verbatim}
\label{ex:singleton}
\end{example}

It is also possible to use named graphs directly (denoted ng in Table~\ref{tab:reicomp}).  A named graph concept allows assigning an IRI for one or more triples as a graph name. 
In that scenario, the graph name is used as a subject that can store the metadata about the associated triples. In \cite{groth2010} authors 
present a model of nanopublications along with a named graph notation.
\begin{example}
The example presents a named graph with metadata.
\begin{verbatim}
<g1> { <#js> foaf:name "John Smith" }
<g1> ex2:certainty 0.5 .
\end{verbatim}
\label{ex:nanopublications}
\end{example}

In Table~\ref{tab:reicomp} we present features of the above-mentioned standardized serializations, namely: having W3C Recommendation, special syntax that is an extension of RDF, and number of extra RDF statements required to represent an RDF triple (i.e. $O(n)$). 

\begin{table}[!t]
\caption[RDF reifications]{RDF reifications. This table presents the features of reifications, namely: having W3C Recommendation (yes -- \CheckedBox, no -- \XBox), RDF-compatible syntax (yes -- \CheckedBox, partial -- \Square, no -- \XBox) and number of extra statements.}
\label{tab:reicomp}
\centering
\begin{tabular}{|l|l|l|l|l|l|}
\hline
\textbf{Feature} & \textbf{sr} & \textbf{nr} & \textbf{rdr} & \textbf{sp} & \textbf{ng} \\
\hline
Standard         & \CheckedBox & \CheckedBox$^\star$ & \XBox        & \XBox       & \CheckedBox \\
RDF syntax only  & \CheckedBox & \CheckedBox & \XBox        & \XBox       & \Square     \\
Extra statements & $4n$        & $2n$        & $n$$^{\star\star}$         & $2n$        & $n$$^{\star\star\star}$          \\
\hline
\multicolumn{6}{l}{$^\star$ W3C Working Group Note}\\
\multicolumn{6}{l}{$^{\star\star}$ Counted as n-tripleX}\\
\multicolumn{6}{l}{$^{\star\star\star}$ Counted as n-quads}\\
\end{tabular}
\end{table}

\section{Modeling Blank Nodes}
\label{sec:bnodes}

The standard semantics for blank nodes interprets them as existential variables. We provide an alternative formulation for the blank nodes, and look at theoretic aspects of blank nodes.

Following \cite{Chen2012}, blank nodes give the capability to:
\begin{itemize}
 \item define the information to encapsulate the N-ary association,
 \item describe reification,
 \item offer protection of the inner data,
 \item describe multi-component structures (e.g. RDF containers),
 \item represent complex attributes without having to name explicitly the auxiliary node.
\end{itemize}

The complexity of the problem of deciding whether or not two RDF graphs with blank nodes are isomorphic is \textbf{GI}-complete as noticed 
in \cite{carroll2002}. Graph isomorphism complexity with a total absence of blank nodes is \textbf{PTIME} \cite{Mallea2011}.

There is a complication in the notion of RDF graphs, caused by blank nodes. Blank nodes are intended to be locally-scoped terms 
that are interpreted as existential variables. Blank nodes are shared by graphs only if they are derived from the ones described by 
documents or RDF datasets that provide for the blank nodes sharing between different graphs. Performing a document download is not 
synonymous with the blank nodes in a resulting graph being identical to the blank nodes coming from the same source or other 
downloads of the same file. This gives rise to a notion of \emph{isomorphism} between RDF graphs that are the same up to blank 
node relabeling: isomorphic RDF graphs can be considered as containing the same content. 

\begin{definition}[Graph isomorphism]
\label{def:gisomorphism}
Two RDF graphs $G_1$ and $G_2$ are identical in form (are \emph{isomorphic}, denoted $G_1 \cong G_2$) if there is a bijective 
function $M$ between the nodes sets of the two graphs, such that:
\begin{enumerate}
\item $M: \mathcal{B} \rightarrow \mathcal{B}$,
\item $M_{|\mathcal{L}} = \mathrm{id_\mathcal{L}}$, for all literals that are nodes of graph $G_1$,
\item $M_{|\mathcal{I}} = \mathrm{id_\mathcal{I}}$, for all RDF IRIs that are nodes of graph $G_1$,
\item $\forall _ {s, p, o} (\langle s, p, o \rangle \in G_1 \Leftrightarrow \langle M(s), p, M(o) \rangle \in G_2)$.
\end{enumerate}\qed
\end{definition}

Moreover, merging two or more RDF graphs is important to ensure that there are no conflicts in blank node labels. A merging operation performs the union after forcing all 
of shared blank nodes that are present in two or more graphs to be distinct in each RDF graph. The graph after this operation is called 
the \emph{merge}. The result of this operation on RDF graphs can produce more nodes than the original graphs.

\begin{definition}[RDF merge]
\label{def:rdfmerg}
Given two graphs, $G_1$ and $G_2$, an \emph{RDF merge} of these two graphs, denoted $G_1 \uplus G_2$, is defined as the set union 
$G_1^\prime \cup G_2^\prime$, where $G_1^\prime$ and $G_2^\prime$ are isomorphic copies of $G_1$ and $G_2$ respectively such that the 
copies do not share any blank nodes.\qed
\end{definition}

As in the RDF graphs we can compare RDF datasets.

\begin{definition}[Dataset isomorphism]
\label{def:disomorphism}
Two RDF datasets $DS_1$ and $DS_2$ are \emph{dataset-isomorphic} if there is a function $M: \mathcal{B} \rightarrow \mathcal{B}$ 
between the triples, graphs and nodes in the two datasets, respectively, such that:
\begin{enumerate}
\item $M$ is the identity map on RDF literals and IRIs, i.e. $M_{|\mathcal{I} \cup \mathcal{L}} = \mathrm{id_{\mathcal{I} \cup \mathcal{L}}}$,
\item for every $G \in DS_1$ such that $G = \{ t_1, t_2, \dots, t_n\}$, $M(G) = \{ M(t_1), M(t_2), \dots, M(t_n) \} \in DS_2$, 
\item for every $G \in DS_1$, $\forall _ {s, p, o} \langle s, p, o \rangle \in G \Leftrightarrow \langle M(s), M(p), M(o) \rangle \in M(G)$,
\item $\forall _ {u, G} (\langle u, G \rangle \in NG_1 \Leftrightarrow \langle M(u), M(G) \rangle \in NG_2)$.
\end{enumerate}\qed
\end{definition}

Blank nodes identifiers cannot be found in the RDF abstract syntax. Giving a constant name to the blank nodes can be helped 
by the skolemization mechanism \cite{hogan2015}. In situations where stronger identification is needed, some or all of the blank nodes can 
be replaced with IRIs. Systems that wish to do so ought to create a globally unique IRI (called a \emph{skolem IRI}) for 
every blank node so replaced. This conversion does not significantly change the graph meaning. It permits the 
possibility of other RDF graphs by subsequent use the skolem IRIs that is impossible for blank nodes. Systems use a 
\emph{well-known IRI} \cite{Nottingham2010} when they need skolem IRIs to be distinguishable outside of the system boundaries 
with the registered name \texttt{genid}.

\begin{definition}[Skolemization]
\label{def:dskolemization}
Assume that $G$ is a graph including blank nodes, $\xi: \mathcal{B} \rightarrow \mathcal{I}_{skolem}$ is a skolemization injective function 
and $\mathcal{I}_{skolem}$ is a subset of skolem IRIs which is substituted for a blank node and not occur in any other RDF graph.\qed
\end{definition}

From the above definition it follows that $\mathcal{I}_{skolem} \cap \mathcal{I}_G = \emptyset$, where $\mathcal{I}_G$ is a set of 
IRIs that are used in $G$.

On the other hand, in \cite{Mallea2011}, authors propose two skolemization schemes, such as: centralized and decentralized.
The first one is very similar to what the URL shortening service does. Formally, there is a distinguished subset of the URIs. 
Whenever the service receives a request, it returns an element that is in Skolem constants such that it has not been used before.
The second proposal resembles the first but with no central service. As a result, each publisher will generate its constants locally.
Another proposal \cite{Hogan2015b} focuses on a scheme to produce canonical labels for blank nodes, which maps them to globally 
canonical IRIs. It guarantees that two skolemized graphs will be equal if and only if the two RDF graphs are isomorphic.

\textbf{NP}-completeness originate from cyclic blank nodes. In \cite{Hogan2014}, authors discuss a number of possible alternatives 
for blank nodes, such as:
\begin{enumerate}
 \item disallow blank node,
 \item ground semantics,
 \item well-behaved RDF.
\end{enumerate}
The first alternative disallows the use of blank nodes in RDF. However, blank nodes are a useful convenience for publishers. The 
second alternative proposes to assign blank nodes a ground semantics, such that they are interpreted in a similar fashion to IRIs. 
The third alternative is also presented in \cite{Boothwell2012}. The core motivation for this proposal is to allow implementers 
to develop tractable lightweight methods that support the semantics of blank nodes for an acyclical case. Following \cite{Boothwell2012},
a \emph{well-behaved RDF graph} is a graph that conforms to the restrictions, which we present below.

\begin{definition}[Well-behaved RDF graph]
\label{def:wellbehaved}
A \emph{well-behaved RDF graph} is an RDF graph that conforms to the following restrictions:
\begin{enumerate}
 \item it can be serialized as Turtle without the use of explicit blank node identifiers,
 \item it uses no deprecated features of RDF.
\end{enumerate}\qed
\end{definition}

Note that the first version of RDF published in 1999 did not have named blank nodes and thus was per definition well-behaved.

Another important concept is leanness, which is checking if an RDF graph contains redundancy.

\begin{definition}[Subgraph and lean graph]
\label{def:leangraph}
A \emph{subgraph} is a subset of a graph $G$. Assume that a function 
$\nu : \mathcal{I} \cup \mathcal{B} \cup \mathcal{L} \rightarrow \mathcal{I} \cup \mathcal{B} \cup \mathcal{L}$
preserving IRIs and literals. An RDF graph $G$ is \emph{lean} if there is no function $\nu$ such that $\nu(G)$ is a 
subgraph of $G$.\qed
\end{definition}

Please note that a subgraph can be a graph with fewer triples. The complexity of the problem of verifying whether or not an RDF 
graph is lean is \textbf{coNP}-complete as noticed in \cite{Gutierrez2011}. Alongside the notion of graphs being non-lean, 
we also intuitively refer to blank nodes as being non-lean. Non-lean blank nodes are the cause of redundant triples in non-lean graphs. 
A graph is non-lean if and only if it contains one or more non-lean blank nodes.

\begin{example}
The example presents that the top graph is lean, because there is no proper map into itself. The bottom graph in not lean.
\begin{center}
\includegraphics{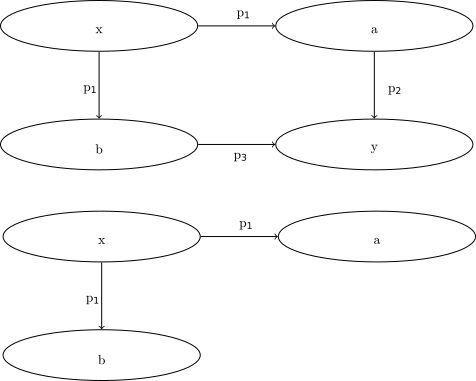}
\end{center}
\label{ex:lean}
\end{example}

\section{Entailments}
\label{sec:entailments}
An interpretation in RDF is a function from literals and IRIs into a set, together with restrictions upon the mapping and the set. 
In this section we introduce different interpretation notions from the RDF area, each corresponding to an entailment regime in a standard way.

Following \cite{Patel-Schneider2014}, a \emph{simple interpretation} $I$ is a structure which we present below.

\begin{definition}[Simple interpretation]
\label{def:sinterpret}
A \emph{simple interpretation} is $I = \langle R^I, P^I, EXT^I, INT^I \rangle$, 
where:
\begin{enumerate}
\item $R^I$ is a (nonempty) set of named resources (the universe of $I$),
\item $P^I$ is a set,  called the set of properties of $I$,
\item $EXT^I$ is an extension function used to associate properties with their property extension,
$EXT^I: P^I \rightarrow 2^{R^I \times R^I}$,
\item $INT^I$ is the interpretation function which assigns a resource or a property to every element of $V$ such that $INT^I$ is 
the identity for literals, $INT^I: V \rightarrow R^I \cup P^I$.
\end{enumerate}\qed
\end{definition}

The interpretation is a map from expressions i.e. triples, graphs and names to truth values and universe elements. Following terminology, 
we say that $I$ \emph{satisfies} an RDF graph $H$ when $I(H)=true$, that $H$ is \emph{satisfiable} if it is a simple interpretation that 
satisfies it. Moreover, an RDF graph $G$ \emph{entails} a graph $H$ when every interpretation that satisfies $G$, will satisfy $H$ as well. 
In this case, we shall write $G \models H$. The graphs $G$ and $H$ are logically \emph{equivalent} if each entails the other. Simple entailment 
can be directly stated in terms of graph homomorphism as noticed in \cite{Patel-Schneider2014}.

Ter Horst also proved the \textbf{NP}-completeness of simple entailment by reduction from the clique problem \cite{Terhorst2005}. If the RDF
graph $H$ lacks of blank nodes, this issue is in \textbf{PTIME}, because one should only check that each triple in 
$H$ is in $G$ too.

Following \cite{Patel-Schneider2014}, blank nodes are seen as simply indicating a thing's existence, without handling an IRI to 
identify any name of a resource. We need to define a version of the simple interpretation mapping that includes the set of blank 
nodes as part of its domain.
\begin{definition}[Semantic condition for blank nodes]
\label{def:condotionbnodes}
Let $G$ be a graph and $I$ be a simple interpretation. Let $\alpha: \mathcal{B} \rightarrow R^I$ be a map from blank nodes to 
resources and let $INT_\alpha$ denote an amended version of $INT^I$ that includes $\mathcal{B}$ as part of its domain such that 
$INT^I(x) = \alpha(x)$ for $x \in \mathcal{B}$ and $INT_\alpha(x) = INT^I(x)$ for $x \in \mathcal{I} \cup \mathcal{L}$. We say that $I$ 
is a model of graph $G$ if there exists a function $\alpha$ such that for each $\langle s, p, o\rangle \in G$, it holds that 
$INT^I(p) \in P^I$ and $(INT_\alpha(s), INT_\alpha(o)) \in EXT^I(INT^I(p))$.\qed
\end{definition}

When two RDF graphs share a blank node, their meaning is not fully captured by treating them in isolation. RDF graphs can be viewed as 
conjunctions of simple atomic sentences in First-Order Logic (FOL)~\cite{date1975}, where blank nodes are free variables which are 
existential.

Further interpretations are depending on which IRIs are known as datatypes. Hayes and Patel-Schneider \cite{Patel-Schneider2014} 
propose the use of a parameter $D$ (the set of known datatypes) on simple interpretations. The next considered interpretation 
is \emph{$D$-interpretation}.

\begin{definition}[$D$-interpretation]
\label{def:dinterpret}
Let $D$ be a set of IRIs describing datatypes. A \emph{$D$-interpretation} is a simple interpretation additionally satisfying the following conditions:
\begin{enumerate}
\item If \texttt{rdf:langString} $ \in D$, then for every language-tagged string $E$ with lexical form $s_l$ and language tag $t_l$, 
$L^I(E) = \langle s_l, t_l^\prime \rangle$, where $t_l^\prime$ is $t_l$ transformed to lower case,
\item For every other IRI $d \in D$, $I(d)$ is the datatype identified by $d$, and for every literal 
"$s_l$"\^{}\^{}$d$, $L^I($"$s_l$"\^{}\^{}$d) = L2V(I(d))(s_l)$, where $L2V$ is a function from datatypes to their lexical-to-value mapping. 
\end{enumerate}\qed
\end{definition}

Note that in the RDF~1.0 specification, datatype $D$-entailment was described as an RDFS-entailment semantic extension. In RDF~1.1 it is 
defined as a simple direct extension. Moreover, in RDF~1.1 datatype entailment formally refers to a set of recognized 
datatypes IRIs. RDF~1.0 used the concept of a datatype map: in the new semantic description, this is the mapping from 
recognized IRIs to the datatypes they identify.

A graph is \emph{satisfiable} recognizing $D$ (or simply $D$-satisfiable) if it has the true in some $D$-interpretation, and $G$ 
\emph{entails} recognizing $D$ (or $D$-entails) $H$ when every $D$-interpretation satisfies $G$, will satisfy satisfies $H$ as well.

In \cite{Terhorst2005} ter Horst proposes the $D*$ semantics, which is a weaker variant of RDFS 1.0 $D$ semantics. This semantics
generalizes the RDFS 1.0 semantics \cite{Hayes2004} to add reasoning with datatypes. 

RDF interpretation imposes additional semantic conditions on part of the (infinite) set of IRIs with the namespace prefix \texttt{rdf:}
and \texttt{xsd:string} datatype. In RDF there are three key terms:
\begin{itemize}
\item \texttt{rdf:Property} ($\mathbf{P}$) -- the class of RDF properties,
\item \texttt{rdf:type} ($\mathbf{a}$) -- the subject is an instance of a class,
\item \texttt{rdf:langString} ($\mathbf{ls}$) -- the class of language-tagged string literal values.
\end{itemize}

\begin{definition}[RDF interpretation]
\label{def:rdfinterpret}
An \emph{RDF interpretation} is a $D$-interpretation $I$ where $D$ has \texttt{xsd:string} and \texttt{rdf:langString}, which satisfies 
the following conditions:
\begin{enumerate}
\item $x \in P^I \Leftrightarrow \langle x, I(\mathbf{P}) \rangle \in EXT^I(\mathbf{a})$,
\item $\forall _ {d \in I} (\langle x, I(d) \rangle \in EXT^I(I(\mathbf{a})) \Leftrightarrow x \in VS(I(d)))$, where $VS$ is a value space.
\end{enumerate}\qed
\end{definition}

This RDF vocabulary is defined by the RDF Semantics \cite{Patel-Schneider2014} in terms of the RDF model theory. A selection of the 
inference rules are presented in Table \ref{tab:rdfrules}. As earlier $G$ RDF \emph{entails} $H$ recognizing $D$ when RDF interpretation 
recognizing $D$ that satisfies $G$, will satisfy $H$ as well. When $D$ is \{\texttt{xsd:string}, \texttt{rdf:langString}\} then $G$ RDF 
entails $H$.

\begin{table}[!t]
\caption[A selection of RDF rules]{A selection of RDF rules. A rule of the form body $\Rightarrow$ head has predicates to the left of the $\Rightarrow$ symbol are the premise of the rule, and predicates to the right of the $\Rightarrow$ symbol are the consequent.}
\label{tab:rdfrules}
\centering
\begin{tabular}{|l|l|l|}
\hline
\bf Rule ID & \bf Body & \bf Head \\
\hline
rdf1 & $?s$ $?p$ "$s_l$"\^{}\^{}$d$ .  $d \in D$ & $\Rightarrow$ $?s$ $?p$ \_:n . \_:n $\mathbf{a}$ $d$ . \\
rdf2 & $?s$ $?p$ $?o$ . & $\Rightarrow$ $?p$ $\mathbf{a}$ $\mathbf{P}$ . \\
\hline
\end{tabular}
\end{table}

RDF Schema \cite{Brickley2014a} extends RDF to additional vocabulary with more complex semantics. RDFS semantics introduces a \emph{class}. 
It is a resource which constitutes a set of things that all have class as a value of their \texttt{rdf:type} predicate
(so-called property). They are outlined to be things of type \texttt{rdfs:Class}. We introduce $C^I$ which is the set of all classes 
in an interpretation. Additionally, the semantic conditions are stated in terms of a function $CEXT^I: C^I \rightarrow 2^{R^I}$. In 
RDFS there are ten key terms:
\begin{itemize}
\item \texttt{rdfs:Class} ($\mathbf{C}$) -- the class of classes,
\item \texttt{rdfs:Literal} ($\mathbf{Lit}$) -- the class of literal values,
\item \texttt{rdfs:Resource} ($\mathbf{Res}$) -- the class resource, everything,
\item \texttt{rdfs:Datatype} ($\mathbf{Dt}$) -- the class of RDF datatypes,
\item \texttt{rdfs:subPropertyOf} ($\mathbf{spo}$) -- the property that allows for stating that all things related by a given property $x$ are also 
necessarily related by another property $y$,
\item \texttt{rdfs:subClassOf} ($\mathbf{sco}$) -- the property that allows for stating that the extension of one class $X$ is necessarily contained 
within the extension of another class $Y$,
\item \texttt{rdfs:domain} ($\mathbf{dom}$) -- the property that allows for stating that the subject of a relation with a given property $x$ is a member of a given class $X$,
\item \texttt{rdfs:range} ($\mathbf{rng}$) -- the property that allows for stating that the object of a relation with a given property $x$ is a member of a given class $X$,
\item \texttt{rdfs:ContainerMembershipProperty} ($\mathbf{Cmp}$) -- the class of container membership properties, \texttt{rdf:\_$i$},
\item \texttt{rdfs:member} ($\mathbf{m}$) -- a member of the subject resource.
\end{itemize}

\begin{definition}[RDFS interpretation]
\label{def:rdfsinterpret}
An \emph{RDFS interpretation} is an RDF interpretation which additionally satisfies the following conditions:
\begin{enumerate}
\item $CEXT^I(y) = \{x: \langle x, y \rangle \in EXT^I(INT^I(\mathbf{a}) \}$,
\item $IC = CEXT^I(INT^I(\mathbf{C}))$,
\item $LV = CEXT^I(INT^I(\mathbf{Lit}))$,
\item $CEXT^I(INT^I(\mathbf{Res})) = R^I$,
\item $CEXT^I(INT^I(\mathbf{ls})) = \{INT^I(E) : E$ is a language-tagged string $\}$,
\item $\forall _ {d \in \mathcal{I}\backslash (\mathbf{Res} \cup \mathbf{ls})} CEXT^I(INT^I(d)) = VS(INT^I(d))$, where $VS$ is a value space,
\item $\forall _ {d \in \mathcal{I}} INT^I(d) \in CEXT^I(INT^I(\mathbf{Dt}))$
\item $\langle x, y \rangle \in EXT^I(INT^I(\mathbf{dom})) \wedge \langle u, v \rangle \in EXT^I(x) \Rightarrow u \in CEXT^I(y)$,
\item $\langle x, y \rangle \in EXT^I(INT^I(\mathbf{rng})) \wedge \langle u, v \rangle \in EXT^I(x) \Rightarrow v \in CEXT^I(y)$,
\item  $\forall _ {x,y,z \in P^I} ( \langle x, y \rangle \in EXT^I(INT^I(\mathbf{spo})) \wedge \langle y, z \rangle \in EXT^I(INT^I(\mathbf{spo})))
\Rightarrow \langle x, z \rangle \in EXT^I(INT^I(\mathbf{spo}))) \wedge \forall _ {x \in P^I} \langle x, x \rangle \in EXT^I(INT^I(\mathbf{spo}))$
\item $\langle x, y \rangle \in EXT^I(INT^I(\mathbf{spo})) \Rightarrow x \in P^I \wedge y \in P^I \wedge CEXT^I(x) \subset CEXT^I(y)$,
\item $x \in C^I \Rightarrow \langle x, INT^I(\mathbf{Res}) \rangle \in EXT^I(INT^I(\mathbf{sco}))$
\item  $\forall _ {x,y,z \in P^I} ( \langle x, y \rangle \in EXT^I(INT^I(\mathbf{sco})) \wedge \langle y, z \rangle \in EXT^I(INT^I(\mathbf{sco})))
\Rightarrow \langle x, z \rangle \in EXT^I(INT^I(\mathbf{sco}))) \wedge \forall _ {x \in P^I} \langle x, x \rangle \quad \in \quad EXT^I(INT^I(\mathbf{sco}))$
\item $\langle x, y \rangle \in EXT^I(INT^I(\mathbf{sco})) \Rightarrow x \in C^I \wedge y \in C^I \wedge CEXT^I(x) \subset CEXT^I(y)$,
\item $x \in CEXT^I(INT^I(\mathbf{Cmp})) \Rightarrow \langle x, INT^I(\mathbf{m}) \rangle \in EXT^I(INT^I(\mathbf{spo}))$,
\item $x \in CEXT^I(INT^I(\mathbf{Dt})) \Rightarrow \langle x, INT^I(\mathbf{Lit}) \rangle \in EXT^I(INT^I(\mathbf{sco}))$.
\end{enumerate}\qed
\end{definition}

This RDFS vocabulary is defined by the RDF Semantics \cite{Patel-Schneider2014} in terms of the RDF model theory. Selections of the RDFS inference 
rules are presented in Table \ref{tab:rdfsrules}. 

As earlier $G$ RDFS entails $H$ recognizing $D$ when every RDFS interpretation recognizing $D$ which satisfies $G$, will satisfy $H$ as well.

\begin{table}[!t]
\caption[A selection of RDFS rules]{A selection of RDFS rules. A rule of the form body $\Rightarrow$ head has predicates to the left of the $\Rightarrow$ symbol are the premise of the rule, and predicates to the right of the $\Rightarrow$ symbol are the consequent.}
\label{tab:rdfsrules}
\centering
\begin{tabular}{|l|l|l|}
\hline
\bf Rule ID & \bf Body & \bf Head \\
\hline
rdfs1 & any IRI $?p \in D$ & $\Rightarrow$ $?p$ $\mathbf{a}$ $\mathbf{Dt}$ . \\
rdfs2 & $?p$ $\mathbf{dom}$ $?x$ . $?y$ $?p$ $?z$ . & $\Rightarrow$ $?y$ $\mathbf{a}$ $?x$ . \\
rdfs3 & $?p$ $\mathbf{rng}$ $?x$ . $?y$ $?p$ $?z$ . & $\Rightarrow$ $?z$ $\mathbf{a}$ $?x$ . \\
rdfs4a & $?x$ $?p$ $?y$ . & $\Rightarrow$ $?x$ $\mathbf{a}$ $\mathbf{Res}$ . \\
rdfs4b & $?x$ $?p$ $?y$ . & $\Rightarrow$ $?y$ $\mathbf{a}$ $\mathbf{Res}$ . \\
rdfs5 & $?x$ $\mathbf{spo}$ $?y$ . $?y$ $\mathbf{spo}$ $?z$ . & $\Rightarrow$ $?x$ $\mathbf{spo}$ $?z$ . \\
rdfs6 & $?x$ $\mathbf{a}$ $\mathbf{P}$ . & $\Rightarrow$ $?x$ $\mathbf{spo}$ $?x$ . \\
rdfs7 & $?p$ $\mathbf{spo}$ $?q$ . $?x$ $?p$ $?y$ . & $\Rightarrow$ $?x$ $?q$ $?y$ . \\
rdfs8 & $?x$ $\mathbf{a}$ $\mathbf{C}$ & $\Rightarrow$ $?x$ $\mathbf{sco}$ $\mathbf{Res}$ . \\
rdfs9 & $?x$ $\mathbf{sco}$ $?y$ . $?z$ $\mathbf{a}$ $?x$ . & $\Rightarrow$ $?z$ $\mathbf{a}$ $?y$ . \\
rdfs10 & $?x$ $\mathbf{a}$ $\mathbf{C}$ . & $\Rightarrow$ $?x$ $\mathbf{sco}$ $?x$ . \\
rdfs11 & $?x$ $\mathbf{sco}$ $?y$ . $?y$ $\mathbf{sco}$ $?z$ . & $\Rightarrow$ $?x$ $\mathbf{sco}$ $?z$ . \\
rdfs12 & $?x$ $\mathbf{a}$ $\mathbf{Cmp}$ . & $\Rightarrow$ $?x$ $\mathbf{spo}$ $\mathbf{m}$ . \\
rdfs13 & $?x$ $\mathbf{a}$ $\mathbf{Dt}$ . & $\Rightarrow$ $?x$ $\mathbf{sco}$ $\mathbf{Lit}$ . \\
\hline
\end{tabular}
\end{table}

\begin{example}
The example presents an RDFS interpretation for the vocabulary presented in Example~\ref{ex:rdfgraphex}.
\begin{center}
\begin{tabular}{rcl}
$R^I \quad = $   & $\{\Upsilon, \Phi, \Psi, \Omega, \alpha, \beta, \gamma, \delta, \epsilon\}$ & \\
$P^I \quad = $   & $\{\Upsilon, \Phi, \Psi, \Omega\}$ & \\
$EXT^I \quad = $ & $\Upsilon$ & $\mapsto \quad \{\langle\alpha, \beta\rangle\}$\\
                 & $\Phi$     & $\mapsto \quad \{\langle\alpha, \delta\rangle\}$\\
                 & $\Psi$     & $\mapsto \quad \{\langle\alpha, \gamma\rangle\}$\\
                 & $\Omega$   & $\mapsto \quad \{\langle\gamma, \epsilon\rangle\}$\\
$INT^I \quad = $ & rdf:type               & $ \mapsto \quad \Upsilon$\\
                 & foaf:name              & $ \mapsto \quad \Phi$\\
                 & foaf:workplaceHomepage & $ \mapsto \quad \Psi$\\
                 & rdfs:label             & $ \mapsto \quad \Omega$\\
                 & \textless \#js\textgreater                 & $ \mapsto \quad \alpha$\\
                 & foaf:Person            & $ \mapsto \quad \beta$\\
                 & "John Smith"           & $ \mapsto \quad \gamma$\\
                 & \textless http://univ.com/\textgreater     & $ \mapsto \quad \delta$\\
                 & "University"           & $ \mapsto \quad \epsilon$
\end{tabular}
\end{center}
\label{ex:interpretation}
\end{example}

In Table \ref{tab:entailmentc} we summarize results about complexity of entailment problems mentioned above.

\begin{table}[!t]
\caption[Entailments complexity]{Computational complexity of entailments for RDF and RDFS}
\label{tab:entailmentc}
\centering
\begin{tabular}{|l|c|c|c|}
\hline
\bf Entailment & \bf Current semantics & \bf No blank nodes \\
\hline
\bf simple & \textbf{NP}-complete  & \textbf{PTIME} \\
\bf D* & \textbf{NP}-complete &  \textbf{PTIME} \\
\bf RDF & \textbf{NP}-complete & \textbf{PTIME} \\
\bf RDFS & \textbf{NP}-complete & \textbf{PTIME} \\
\hline
\end{tabular}
\end{table}

\begin{example}
The example presents that the top graph entails the bottom graph. When we blank the node \texttt{<\#js>} from the top, the bottom graph still 
has a node that represents a person and preserves semantics. Similarly, when we delete the node \texttt{"John Smith"} from the top graph, the
bottom graph still preserves the graph's semantics.
\begin{center}
\includegraphics{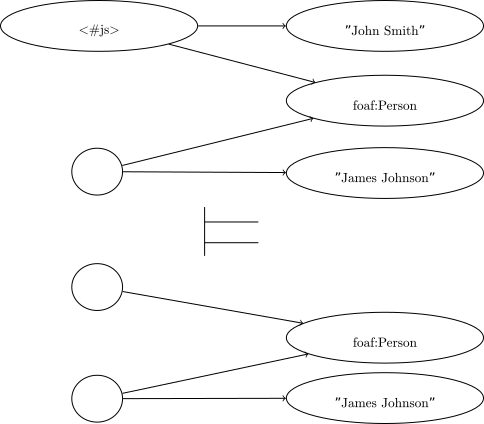}
\end{center}
\label{ex:?}
\end{example}

\section{RDF Data Integration}
\label{sec:integration}
The role of RDF as an integration system for data from different sources is one of the crucial motivations for research efforts. It is important 
to provide integration methods to bridge the gap between the RDF and other environments. Following \cite{lenzerini2002}, the definition of data 
integration systems is provided below.

\begin{definition}[Data integration system]
\label{def:dataintegration}
A \emph{data integration system} is a tuple $\langle \mathcal{G}, S, \mathcal{M}\rangle$ where $\mathcal{G}$ is the global schema, $S$ is the 
source schema and $\mathcal{M}$ is the mapping between $\mathcal{G}$ and $S$, constituted by a set of assertions.\qed
\end{definition}

In this section the approaches of mapping of relational databases to RDF are discussed (Subsection \ref{sec:rdb2rdf}) as well as approaches of 
mapping XML to RDF (Subsection \ref{sec:xml2rdf}). However, there are also more general 
approaches \cite{corby2015a,corby2015b,alkhateeb2008,quan2005,peroni2015,Tandy2015,dimou2014}.

\subsection{Bringing Relational Databases into the RDF}
\label{sec:rdb2rdf}
This subsection contains an overview and comparison of the approaches for mapping from a relational database into RDF. Table \ref{tab:relational} presents 
the key approaches from related work. It presents the features of the below-mentioned proposals, namely: mapping representation (SQL, RDF, 
XML, etc.), schema representation (RDF, OWL \cite{patel-schneider2002} and F-Logic \cite{kifer1989}) and level of automation.

\begin{table}[!t]
\caption[Relational databases mapping]{Relational databases mapping. This table presents the features of transformation approaches, namely: mapping representation, schema representation and level of automation (automatic -- \CheckedBox, semi-automatic -- \Square\: and manual -- \XBox)}.
\label{tab:relational}
\centering
\begin{tabular}{|c|l|l|c|}
\hline
\bf Approaches                 & \bf Mapping                  & \bf Schema                  & \bf Automatic        \\
\                              & \bf Represent.               & \bf Represent.              &                      \\
\hline
\cite{astrova2004,astrova2007} & n/a                          & RDFS, OWL, F-Logic          & \Square              \\
\cite{auer2009}                & SQL                          & RDFS                        & \XBox                \\
\cite{cerbah2008a,cerbah2008b} & Constraint rules             & RDFS, OWL                   & \Square              \\
\cite{bizer2004}               & D2RQ                         & RDFS                        & \CheckedBox          \\
\cite{curino2009}              & RDF, Rel.OWL                 & RDFS, OWL                   & \CheckedBox          \\
\cite{buccella2004}            & n/a                          & RDFS, OWL                   & \Square              \\
\cite{hu2007}                  & FOL, Horn                    & RDFS, OWL                   & \CheckedBox          \\
\cite{bumans2010}              & SQL                          & RDFS, OWL                   & \XBox                \\
\cite{li2005}                  & Logic rules                  & RDFS, OWL                   & \Square              \\
\cite{byrne2008}               & n/a                          & RDFS                        & \Square              \\
\cite{calvanese2011}           & XML                          & RDFS, OWL                   & \XBox                \\
\cite{erling2007}              & SPARQL                       & RDFS                        & \CheckedBox$^\star$  \\
\cite{ghawi2007}               & R$_2$O                       & RDFS, OWL                   & \CheckedBox          \\
\cite{hert2010}                & RDF                          & RDFS                        & \XBox                \\
\cite{nyulas2007}              & RDF, Rel.OWL                 & RDFS, OWL                   & \CheckedBox          \\
\cite{polflietI2010}           & D2RQ                         & RDFS, OWL                   & \Square              \\
\cite{sahoo2008}               & XPath (XSLT)                 & RDFS, OWL                   & \XBox                \\
\cite{salas2010}               & SQL                          & RDFS, OWL                   & \Square              \\
\cite{shen2006}                & n/a                          & RDFS, OWL                   & \CheckedBox          \\
\cite{stojanovic2002}          & n/a                          & RDFS, F-Logic               & \Square              \\
\cite{tirmizi2008}             & FOL                          & RDFS, OWL                   & \CheckedBox          \\
\cite{vavliakis2013}           & SQL                          & RDFS, OWL                   & \XBox                \\
\cite{wu2006}                  & RDF/XML                      & RDFS, OWL                   & \XBox                \\
\cite{seaborne2007}            & RDF (Direct)                 & RDFS                        & \CheckedBox$^\star$  \\
\cite{lopes2011,bischof2011}   & XQuery                       & n/a                         & \Square              \\
\hline
\multicolumn{4}{r}{$^\star$ possible manual level}\\
\end{tabular}
\end{table}

At the beginning, we focus on solutions \cite{auer2009,salas2010,vavliakis2013} based on SQL as mapping representation. Triplify \cite{auer2009} is 
based on the mapping of HTTP requests onto database queries expressed in SQL queries, which are used to match subsets of the store contents and map 
them to classes and properties. It converts the resulting relations into RDF triples and subsequently publishes it in various RDF syntaxes. 
That proposal includes an approach for publishing update logs in RDF which contain all RDF resources in order to enable incremental crawling of data sources. 
An additional advantage is that it can be easily integrated and deployed with the numerous, widely installed Web applications. The next approach is 
StdTrip \cite{salas2010}, which proposes a structure-based framework using existing ontology alignment software. The approach finds ontology 
mappings between simple vocabulary that is generated from a database. The results of the ontology alignment algorithm are presented as suggestions 
to the user, who matches the most appropriate ontology mapping. RDOTE \cite{vavliakis2013} also uses SQL for the specification of the data subset. 
In that proposal, the suitable SQL query is stored in a file. That approach transforms data residing in the database into RDF graph dump using classes and properties.

The next group of approaches \cite{bizer2004,polflietI2010} uses D2RQ as mapping representation. D2RQ \cite{bizer2004} supports both automatic and 
manual operation modes. In the first mode, RDFS vocabulary is created, in accordance with the reverse engineering methodologies, for the translation 
of foreign keys to properties. In the second mode, the contents of the database are exported to an RDF in accordance with mappings stored in RDF. 
It allows RDF applications to treat non-RDF stores as a virtual RDF graph. It allows RDF applications to treat non-RDF stores as virtual RDF graphs. 
One of the disadvantages of D2RQ is a read-only RDF view of the database. Another D2RQ-based proposal is AuReLi \cite{polflietI2010}, which uses 
several string similarity measures associating attribute names to existing vocabulary entities in order to complete the automation of the 
transformation of databases. It also tries to link database values with ontology individuals. RDF Views \cite{erling2007} has similar functionality 
to D2RQ. It supports both automatic and manual operation modes. That solution uses the table as RDFS class and column as a predicate and takes into 
account cases such as whether a column is part of the unique key. The data is represented as virtual RDF graphs without the physical formation of RDF 
datasets.

Another group of proposals \cite{hert2010,seaborne2007} use RDF. The first approach is OntoAccess \cite{hert2010}, which is a vocabulary-based write 
access to data. That paper consists of the relational database to RDF mapping language called R3M, which consists of an RDF format and 
algorithms for translating queries to SQL. The next proposal is SquirrelRDF \cite{seaborne2007}, which extracts data from a number of databases 
and integrates that data into a business process. That proposal supports RDF views and allows for the execution of queries against it. In that 
group we can also distinguish solutions \cite{curino2009,nyulas2007} based on Relational.OWL \cite{delaborda2005a}. ROSEX \cite{curino2009} uses 
Relational.OWL to represent the relational schema of a database as an OWL ontology. The created database schema annotation and documentation is 
mapped to a domain-specific vocabulary, which is achieved automatically by reverse-engineering the schema. An additional advantage is that it 
supports automatic query translation. DataMaster \cite{nyulas2007} also uses Relational.OWL for the importing of schema structure and data from 
relational databases. 

R2RML~\cite{das2012} is the best-known language based of RDF for expressing mappings from relational databases to RDF 
datasets because it is a W3C recommendation and has a lot of implementations \cite{priyatna2015,sequeda2015,erling2009}. R2RML is a language 
for specifying mappings from relational to RDF data. A mapping takes as input a logical table (\texttt{logicalTable} predicate), i.e. a database 
table, an SQL query, or a database view. In the next step, a logical table is mapped to triples map that is a set of triples. Triples map has two 
main parts. The first part is a subject map (\texttt{subjectMap} predicate) that generates the subject of all RDF triples that will be generated 
from a logical table row. The second part is a predicate-object map (\texttt{predicateObjectMap} predicate) specifies the target property and 
the generation of the object via \texttt{objectMap}.

Yet another type of approaches are \cite{wu2006,calvanese2011}. DartGrid \cite{wu2006} describes a database integration architecture. It uses a 
visual mapping system to align the relational database to existing vocabulary. Correspondences between components of the models which are defined 
via a graphical user interface are created and stored in an RDF/XML syntax (see Subsection \ref{sec:single}). The next proposal is 
MASTRO \cite{calvanese2011}, which is a framework that enables the definition of mappings between a relational database and vocabulary.

Another group of proposals \cite{tirmizi2008,hu2007} use First-Order Logic (FOL) \cite{date1975}. Tirmizi \textit{et al.} \cite{tirmizi2008} present formal 
rules in FOL to transform column to predicate and table to class. The authors present the system which is complete with respect to a space of 
the possible primary key and foreign key mixtures. The next proposal is MARSON \cite{hu2007}, which uses mappings based on virtual documents (called 
vectors of coefficients). It removes incorrect mappings via validating mapping consistency. Moreover, a special type of semantic mappings (called 
contextual mappings) is presented in the paper.

Some approaches \cite{cerbah2008a,ghawi2007,li2005,sahoo2008,tirmizi2008,lopes2011} adopt other mapping representations. RDBToOnto \cite{cerbah2008a,cerbah2008b} 
is a tool that simplifies the implementation of methods for vocabulary acquisition from databases. It provides a visual interface for manual 
modification and adjustment of the learning parameters. RDBToOnto links the data analysis with heuristic rules and generates an ontology. DB2OWL \cite{ghawi2007} 
creates local vocabulary from a relational database that is aligned to reference vocabulary. The vocabulary generated in that proposal reflects the 
database semantics. The mappings are stored in an R$_2$O \cite{barrasa2004} document. The next proposal is SOAM \cite{li2005}, which uses the column to 
predicate and table to class approach. It creates an initial schema, which is refined by referring to a dictionary. Constraints are mapped to 
constraints in the vocabulary schema. That approach tries to establish the quality of the constructed vocabulary. Another proposal is \cite{sahoo2008}. 
It is a domain semantics-driven mapping generation approach. Mapping is created on the XSLT \cite{kay2007} and XPath \cite{snelson2014}. Yet another
approach is XSPARQL \cite{lopes2011}, which can be used both in relational databases and XML (see Subsection~\ref{sec:xml2rdf}).

There are also several approaches \cite{astrova2004,buccella2004,shen2006,stojanovic2002,byrne2008} that do not have a defined mapping representation.
Astrova \cite{astrova2004,astrova2007} discusses correlation among key, data in key attributes between two relations and non-key attributes. In these 
papers, the quality of transformation is considered. Buccella \textit{et al.} \cite{buccella2004}, Shen \textit{et al.} \cite{shen2006} and 
Stojanovic \textit{et al.} \cite{stojanovic2002} examine heuristic rules. Byrne \cite{byrne2008} proposes a domain-specific approach for the generic 
design of cultural heritage data and discusses the options for including published heritage thesauri.

\subsection{Bringing XML into the RDF}
\label{sec:xml2rdf}
This subsection overviews and compares the approaches for mapping from XML into RDF. Table \ref{tab:xml},
presents key approaches from related work. It presents features of the below-mentioned proposals, namely: existing vocabulary, schema representation (RDF, OWL and DAML+OIL \cite{patel-schneider2002}) and level of automation.

\begin{table}[!t]
\caption[XML mapping]{XML mapping. This table presents the features of transformation approaches, namely: existing vocabulary (yes -- \CheckedBox, no -- \XBox), schema representation and level of automation (automatic -- \CheckedBox, semi-automatic -- \Square\: and manual -- \XBox)}
\label{tab:xml}
\centering
\begin{tabular}{|c|c|l|c|}
\hline
\bf Approaches              & \bf Existing              & \bf Schema                  & \bf Auto-          \\
                            & \bf vocabulary            & \bf Representation          & \bf matic          \\
\hline
\cite{amann2002}               & \XBox                     & RDFS, DAML+OIL              & \XBox           \\
\cite{battle2006}              & \XBox                     & RDFS, OWL                   & \CheckedBox     \\
\cite{bedini2011}              & \XBox                     & RDFS, OWL                   & \CheckedBox     \\
\cite{berrueta2008}            & \XBox                     & n/a                         & \CheckedBox     \\
\cite{bikakis2009}             & \XBox                     & RDFS, OWL                   & \CheckedBox     \\
\cite{bischof2011,bischof2012} & \XBox                     & n/a                         & \Square         \\
\cite{bohring2005}             & \XBox                     & RDFS, OWL                   & \CheckedBox     \\
\cite{connolly2007}            & \XBox                     & RDFS, OWL                   & \Square         \\
\cite{cruz2008}                & \CheckedBox               & RDFS, OWL                   & \CheckedBox     \\
\cite{deursen2008}             & \CheckedBox               & RDFS, OWL                   & \Square         \\
\cite{dimou2014}               & \XBox                     & RDFS, OWL                   & \Square         \\
\cite{droop2009}               & \XBox                     & n/a                         & \Square         \\
\cite{farrell2007}             & \XBox                     & n/a                         & \Square         \\
\cite{ferdinand2004}           & \XBox                     & RDFS, OWL                   & \CheckedBox     \\
\cite{garcia2005}              & \XBox                     & RDFS, OWL                   & \CheckedBox     \\
\cite{ghawi2009}               & \CheckedBox               & RDFS, OWL                   & \CheckedBox     \\
\cite{klein2002}               & \XBox                     & RDFS                        & \CheckedBox     \\
\cite{koffina2006}             & \XBox                     & RDFS                        & \CheckedBox     \\
\cite{lehti2004}               & \XBox                     & RDFS, OWL                   & \CheckedBox     \\
\cite{oconnor2011}             & \XBox                     & RDFS, OWL                   & \Square         \\
\cite{reif2004}                & \CheckedBox               & RDFS, OWL                   & \CheckedBox     \\
\cite{rodrigues2008}           & \CheckedBox               & RDFS, OWL                   & \CheckedBox     \\
\cite{shapkin2015}             & \XBox                     & RDFS, OWL                   & \CheckedBox     \\ 
\cite{stavrakantonakis2010}    & \XBox                     & RDFS, OWL                   & \CheckedBox     \\
\cite{thuy2009,thuy2013}       & \XBox                     & RDFS, OWL                   & \Square         \\
\cite{tsinaraki2007}           & \XBox                     & n/a                         & \CheckedBox     \\
\cite{xiao2006}                & \XBox                     & RDFS                        & \CheckedBox     \\
\cite{yahia2012}               & \XBox                     & RDFS, OWL                   & \CheckedBox     \\
\hline
\end{tabular}
\end{table}

At the beginning, we focus on solutions \cite{cruz2008,deursen2008,ghawi2009,reif2004,rodrigues2008} that use existing 
vocabulary and/or ontology. This means that the XML data is transformed according to the mapped vocabularies. 
Cruz \textit{et al.} \cite{cruz2008} propose basic mapping rules to specify the transformation rules on properties, which 
are defined in the XML Schema. Deursen \textit{et al.} \cite{deursen2008} propose the method for the transformation of XML data into RDF 
instances in an ontology-dependent way. X2OWL \cite{ghawi2009} is a tool that builds an OWL ontology from an XML data 
source and a set of mapping bridges. That proposal is based on an XML Schema that can be modeled using different styles to 
create the vocabulary structure. The next proposal is WEESA \cite{reif2004}, which is an approach for Web engineering techniques 
and developing semantically tagged applications. Another tool is JXML2OWL \cite{rodrigues2008}. 
It supports the transformation from syntactic data sources in XML format to a common shared global model defined by vocabulary.

Another group of proposals 
\cite{amann2002,bedini2011,bikakis2009,ferdinand2004,garcia2005,klein2002,koffina2006,lehti2004,oconnor2011,thuy2009,xiao2006,yahia2012}
do not support mappings between XML Schemas and existing vocabularies. Amann \textit{et al.} \cite{amann2002} discuss a data integration 
system, where XML is mapped into vocabulary that supports roles and inheritance. That tool focuses on offering the appropriate 
high-level primitives and mechanisms for representing the semantics of XML data. Janus \cite{bedini2011} is a framework that 
focuses on an advanced logical representation of XML Schema components and a set of patterns that enable the transformation 
from XML Schema into the vocabulary. It supports a set of patterns that enable the translation from XML Schema into ontology. 
SPARQL2XQuery \cite{bikakis2009} is a framework that transforms SPARQL \cite{harris2013} query into a XQuery \cite{robie2014} 
using mapping from vocabulary to XML Schema. It allows query XML databases. Ferdinand \textit{et al.} \cite{ferdinand2004} propose two 
independent mappings: from XML to RDF graphs and XML Schema to OWL. That proposal allows items in XML documents to be mapped 
to different items in OWL. Garcia \textit{et al.} \cite{garcia2005} present a domain-specific approach that maps the MPEG-7 standard 
to RDF. Klein \cite{klein2002} proposes a procedure to transform the XML tree using the RDF primitives by annotating the XML by RDFS. 
This procedure can multiply the availability of semantically annotated RDF data. SWIM \cite{koffina2006} is an integration 
middleware for mediating high-level queries to XML sources using RDFS. Lehti \textit{et al.} \cite{lehti2004} show how the ontologies 
can be used for mapping data sources to a global schema. In this work, the authors introduce how the inference rules can be used to 
check the consistency of such mappings. In this paper, a query language based on XQuery is presented. O'Connor \textit{et al.} \cite{oconnor2011} 
propose an OWL-based language that can transform XML documents to arbitrary ontologies. It extends it with Manchester syntax for 
XPath to support references to XML fragments. Another framework is DTD2OWL \cite{thuy2009}, which changes XML into vocabularies. 
It also allows transforming specific XML instances into OWL individuals. Xiao \textit{et al.} \cite{xiao2006} propose the mappings between 
XML schemas and local RDFS vocabularies and those between local vocabulary and the global RDFS vocabulary. The authors discuss the 
problem of query containment and present a query rewriting algorithm for RDQL \cite{delaborda2005b} and XQuery. Yahi \textit{et al.} \cite{yahia2012} 
propose an approach which covers schema level and data level. In this proposal XML Schema documents are generated for XML documents 
with no schema using the \emph{trang} tool.

Solutions \cite{berrueta2008,bohring2005,droop2009} using XSLT are separate from the ones mentioned above. Berrueta \textit{et al.} \cite{berrueta2008} 
discuss an XSLT+SPARQL framework, which allows to perform SPARQL queries from XSLT. It is a collection of functions for XSLT, which 
allows to transform the XML results format. Yet another tool is XML2OWL \cite{bohring2005}, which uses XSLT for mapping from XML to 
ontology. Droop \textit{et al.} \cite{droop2009} propose another XSLT solution, which allows embedding XPath into SPARQL.
Shapkin \textit{et al.} \cite{shapkin2015} propose a transformation language, which is not strictly based on XLST but inspired by it. 
That proposal focuses on matching of types of RDF resources.

Another subgroup of approaches \cite{bischof2011,battle2006,connolly2007,farrell2007,stavrakantonakis2010,tsinaraki2007} supports mutual 
transformation. XSPARQL \cite{bischof2011,bischof2012} is a query language based on SPARQL and XQuery for transformations from RDF 
into XML and back. It is built on top of XQuery in a syntactic and semantic view. Gloze \cite{battle2006} is another tool for 
bidirectional mapping between XML and RDF. It uses information available in the XML schema for describing how XML is mapped into 
RDF and back again. GRDDL \cite{connolly2007} is a markup language that obtains RDF data from XML documents. It is represented in XSLT. 
SAWSDL \cite{farrell2007} is also a markup language but it proposes a collection of new attributes for the WSDL \cite{chinnici2007} and 
XML Schema. Yet other tools are SPARQL2XQuery \cite{stavrakantonakis2010} and XS2OWL \cite{tsinaraki2007}.

\section{RDF Serializations}
\label{sec:serializations}
Several of RDF syntax formats exist for writing down graphs. RDF~1.1 introduces a number of serialization formats, such as:
Turtle, N-Triples, TriG, N-Quads, JSON-LD, RDFa, and RDF/XML. Note that in RDF~1.1 RDF/XML is no longer the only recommended serialization format.

In Subsection \ref{sec:single}, we present serializations that support single graphs. In Subsection \ref{sec:multiple}, we briefly introduce 
serialization which supports multiple graphs. Moreover, in this section we show the above-mentioned formats in examples.

\subsection{Single Graph Support}
\label{sec:single}
\underline{RDFa} \cite{McCarron2015} (denoted rdfa in Table~\ref{tab:rdfserializations}) is an RDF syntax which embeds RDF triples in HTML and 
XML documents. The RDF data is mixed within Document Object Model. This implies that document content can be marked up with 
RDFa. It adds a set of attribute-level extensions to HTML and various XML-based document types for embedding rich metadata within 
Web documents. What is more, RDFa allows for free intermixing of terms from multiple vocabularies. It is also designed in such a 
way that the format can be processed without information of the specific vocabulary being used. It is common in contexts where data 
publishers are able to change Web templates but have little additional control over the publishing infrastructure.

Following \cite{McCarron2015}, we provide the most important attributes that can be used in RDFa, such as:
\begin{itemize}
\item \texttt{about} -- an attribute that is an IRI or CURIE \cite{Birbeck2010} specifying the resource the metadata is about (an RDF subject),
\item \texttt{rel} and \texttt{rev} -- attributes that expresses (reverse) relationships between two resources (an RDF predicate),
\item \texttt{property} -- an attribute that expresses relationships between a subject and some literal value (an RDF predicate),
\item \texttt{resource} -- an attribute for expressing a relationship's partner resource that is not intended to be navigable (an RDF object),
\item \texttt{href} -- an attribute that expresses the partner resource of a relationship (an RDF resource object),
\item \texttt{src} -- an attribute that expresses a relationship's partner resource when the resource is embedded (an RDF object that is a resource),
\item \texttt{content} -- an attribute that overrides the content of the element when using the \texttt{property} (an RDF object that is a literal),
\item \texttt{datatype} -- an attribute that specifies the datatype of a literal,
\item \texttt{typeof} -- an attribute that specifies the RDF types of the subject or the partner resource,
\item \texttt{inlist} -- an attribute that specifies that the object associated with \texttt{property} or \texttt{rel} attributes on the same 
element is to be pushed onto the list for that predicate,
\item \texttt{vocab} -- an attribute that specifies the mapping to be used when an RDF term is assigned in a value of attribute.
\end{itemize}

\begin{example}
The example presents an RDFa 1.1 serialization that represents RDF triples of Example~\ref{ex:rdfgraphex}.
\begin{verbatim}
<div xmlns="http://www.w3.org/1999/xhtml"
  prefix="
    foaf: http://xmlns.com/foaf/0.1/
    rdf: http://www.w3.org/1999/02/22-rdf-syntax-ns#
    rdfs: http://www.w3.org/2000/01/rdf-schema#"
  >
  <div typeof="foaf:Person" about="http://example.com/p#js">
    <div property="foaf:name">John Smith</div>
    <div rel="foaf:workplaceHomepage">
      <a typeof="rdfs:Resource" href="http://univ.com/">
        <span property="rdfs:label">University</span>
      </a>
    </div>
  </div>
</div>
\end{verbatim}
\label{ex:rdfa}

\end{example}

\underline{RDF/XML} \cite{Gandon2014} (denoted xml in Table~\ref{tab:rdfserializations}) is a syntax to serialize an RDF graph as an XML document. 
Nevertheless, the syntax is viewed as problematic to read and write for humans so one should consider using other syntaxes for editors. In order to 
process a graph in RDF/XML serialization, there must be a representation of nodes and predicates in the terms of Extensible Markup Language 
(XML) -- names of elements, names of attributes, contents of elements and values of attributes. This syntax uses qualified names 
(so-called QNames) to represent IRI references. There are certain restrictions imposed on this serialization by the XML format and the 
use of the XML namespace, which prevents the storage of all RDF graphs as some IRI references are prohibited by the specifications of 
these standards.

Following \cite{Gandon2014}, we provide the most important elements and attributes that can be used in RDF/XML, such as:
\begin{itemize}
\item \texttt{rdf:RDF} -- a root element of RDF/XML documents,
\item \texttt{rdf:Description} -- an element that contains elements that describe the resource, it contains the description of the resource 
identified by the \texttt{rdf:about} attribute,
\item \texttt{rdf:Alt}, \texttt{rdf:Bag} and \texttt{rdf:Seq} -- elements that are containers used to describe a group of things (see Section \ref{sec:preliminaries}),
\item \texttt{rdf:parseType="Collection"} -- an attribute that describe groups that can only contain the specified members,
\item \texttt{rdf:parseType="Resource"} -- an attribute that is used to omit blank nodes,
\item \texttt{xml:lang} -- an attribute that is used to allow content language identification,
\item \texttt{rdf:datatype} -- an attribute that is used to define a typed literal,
\item \texttt{rdf:nodeID} -- an attribute that identify a blank node,
\item \texttt{rdf:ID} and \texttt{xml:base} -- attributes that abbreviate IRIs.
\end{itemize}
\begin{example}
The example presents an RDF/XML serialization that represents RDF triples of Example~\ref{ex:rdfgraphex}.
\begin{verbatim}
<?xml version="1.0" encoding="UTF-8"?>
<rdf:RDF 
   xmlns:foaf="http://xmlns.com/foaf/0.1/"
   xmlns:rdf="http://www.w3.org/1999/02/22-rdf-syntax-ns#"
   xmlns:rdfs="http://www.w3.org/2000/01/rdf-schema#"
>
  <rdf:Description rdf:about="http://univ.com/">
    <rdfs:label>University</rdfs:label>
  </rdf:Description>
  <rdf:Description rdf:about="http://example.com/p#js">
    <foaf:name>John Smith</foaf:name>
    <rdf:type rdf:resource="http://xmlns.com/foaf/0.1/Person"/>
    <foaf:workplaceHomepage rdf:resource="http://univ.com/"/>
  </rdf:Description>
</rdf:RDF>
\end{verbatim}
\label{ex:xmlrdf}
\end{example}

Another RDF syntax refers to Terse RDF Triple Language (\underline{Turtle}) \cite{Prud'hommeaux2014} (denoted ttl in Table~\ref{tab:rdfserializations}). 
That solution offers textual syntax that enables recording RDF graphs in a compact form, including abbreviations that use data patterns and datatypes. 
Following \cite{Prud'hommeaux2014}, we provide the most important rules for constructing the Turtle document:
\begin{itemize}
\item The simplest triple statement consists of a sequence of subject, predicate, and object, separated by space, tabulation or other 
whitespace and terminated by a dot after each triple.
\item Often, the same subject will be referenced by several predicates. In this situation, a series of predicates and objects are separated 
by a semicolon.
\item As with predicates, objects are often repeated with the same subject and predicate. In this case, a comma should be used as a separator.
\item IRIs may be written as relative or absolute IRIs or prefixed names. Both absolute and relative IRIs are enclosed in less-than sign and 
greater-than sign.
\item Quoted Literals have a lexical form followed by a datatype IRI, a language tag or neither. Literals should be delimited by apostrophe 
or double quotes.
\item Blank nodes are expressed as underscore, colon and a blank node label that is a series of name characters. Blank nodes 
can be nested, abbreviated, and delimited by square brackets.
\item Collections are enclosed by parentheses.
\end{itemize}
\begin{example}
The example presents a Turtle serialization that represents RDF triples of Example~\ref{ex:rdfgraphex}.
\begin{verbatim}
@prefix foaf: <http://xmlns.com/foaf/0.1/> .
prefix rdfs: <http://www.w3.org/2000/01/rdf-schema#> .
<http://example.com/p#js> a foaf:Person ;
                          foaf:name "John Smith" ;
                          foaf:workplaceHomepage <http://univ.com/> .
<http://univ.com/> rdfs:label "University" .
\end{verbatim}
\label{ex:turtle}
\end{example}

\underline{N-Triples} \cite{Carothers2014a} (denoted nt in Table~\ref{tab:rdfserializations}) is a line-based, plain text serialization format 
and a subset of the Turtle format minus features such as shorthands. It means that there is a lot of redundancy, and its files can be larger 
than Turtle and RDF/XML. N-Triples was designed to be a simpler format than Turtle, and therefore easier for software to parse and generate. 
Following \cite{Carothers2014a}, we provide the most important rules for constructing the N-Triples document:
\begin{itemize}
\item The triple statement consists of a sequence of subject, predicate, and object, divided by whitespace, and terminated by a dot after each 
triple.
\item IRIs should be represented as absolute IRIs and they are enclosed in less-than sign and greater-than sign.
\item The representation of the lexical form is a sequence of a double quote (an initial delimiter), a list of characters or escape sequence, 
and a double quote (a final delimiter).
\item Blank nodes are expressed as underscore, colon and a blank node label that is a series of name characters.
\end{itemize}
There are some changes in RDF~1.1 N-Triples, e.g. encoding is UTF-8 rather than US-ASCII and blank node labels may begin with a digit.
Syntactically, N-Triples is a subset of Turtle.
\begin{example}
The example presents a N-Triples serialization that represents RDF triples of Example~\ref{ex:rdfgraphex}. Note that we change some RDF 
term because of legibility.
\begin{verbatim}
<http://example.com/p#js> <http://...#type> <http://...foaf/0.1/Person> .
<http://example.com/p#js> <http://.../workplaceH...>  <http://univ.com/> .
<http://example.com/p#js> <http://.../name> "John Smith" .
<http://univ.com/> <http://...#label> "University" .
\end{verbatim}
\label{ex:ntriples}
\end{example}

\subsection{Multiple Graphs Support}
\label{sec:multiple}
\underline{JSON-LD} \cite{Sporny2014} (denoted jld in Table~\ref{tab:rdfserializations}) is a JSON-based format to serialize structured data such 
as RDF. The syntax is designed to easily integrate into deployed systems that use JSON and provides a smooth upgrade path from JSON to JSON-LD. The 
use of RDF in JSON makes RDF data accessible to Web developers without the obligation to install additional parsers, software libraries or other programs 
for changing RDF data. Like JSON, JSON-LD uses human-readable text to transmit data objects consisting of key-value pairs.

Keywords in JSON-LD are starting with at sign. Following \cite{Sporny2014}, we provide the most important keywords that can be used in JSON-LD, such as:
\begin{itemize}
\item \texttt{@context} -- set the short-hand names that are used throughout a document,
\item \texttt{@id} -- uniquely identify things that are being described in the document with blank nodes or IRIs,
\item \texttt{@value} -- specify the data that is associated with a particular property,
\item \texttt{@language} -- define the language for a particular string value or the default language of a document,
\item \texttt{@type} -- set the data type of an IRI, a blank node, a JSON-LD value or a list,
\item \texttt{@container} -- set the default container type for a short-hand string that expands to an IRI or a blank node identifier,
\item \texttt{@list} -- define an ordered set of data,
\item \texttt{@set} -- define an unordered set of data (values are represented as arrays),
\item \texttt{@reverse} -- used for reverse relationship expression between two resources,
\item \texttt{@index} -- specify that a container is used to index information,
\item \texttt{@base} -- define the base IRI against which relative IRIs are resolved,
\item \texttt{@vocab} -- expand properties and values in \texttt{@type} with a common prefix IRI,
\item \texttt{@graph} -- express a graph.
\end{itemize}
\begin{example}
The example presents a JSON-LD serialization that represents RDF triples of Example~\ref{ex:namedgraphex1}.
\begin{verbatim}
{ "@id": "http://example.com/#people",
  "@graph": [
    {
      "@id": "http://univ.com/",
      "http://www.w3.org/2000/01/rdf-schema#label": "University"
    }, 
    {
      "@id": "http://example.com/p#js",
      "@type": "http://xmlns.com/foaf/0.1/Person",
      "http://xmlns.com/foaf/0.1/name": "John Smith",
      "http://xmlns.com/foaf/0.1/workplaceHomepage": {
        "@id": "http://univ.com/" 
       }
    }
  ]
}
\end{verbatim}
\label{ex:jsonld1}
\end{example}

JSON-LD is an RDF syntax. However, it also extends the RDF data model, e.g. in JSON-LD predicates can be IRIs 
or blank nodes whereas in RDF have to be IRIs. JSON-LD can serialize generalized RDF triples, where subjects, predicates, and objects can be 
IRIs, blank nodes or literals.
\begin{example}
The example presents a JSON-LD serialization that represents an RDF triple: \texttt{<\#a>} as a subject, blank
node as a predicate and \texttt{"Alice"} as an object.
\begin{verbatim}
{
 "@context": {
 "name": "_:b"
 },
 "@id": "#a",
 "name": "Alice"
}
\end{verbatim}
\label{ex:jsonld2}
\end{example}

Another RDF syntax refers to \underline{TriG} \cite{Seaborne2014} (denoted trig in Table~\ref{tab:rdfserializations}). It is a plain text syntax 
for RDF datasets serialization. Syntactically, Turtle is subset of TriG. A document consists of:
\begin{enumerate}
\item a sequence of directives,
\item RDF triples,
\item graph statements which contain triple-generating statements.
\end{enumerate}
Graph statements are a pair of blank node label or an IRI with a group of RDF triples surrounded by curly brackets. The blank node label or IRI 
of the graph statement may be used in another graph statement which implies taking the union of the triples generated by each graph statement.
A blank node label or IRI used as a graph label may also reoccur as part of any RDF triples.
\begin{example}
The example presents a TriG serialization that represents RDF triples of Example~\ref{ex:namedgraphex1}.
\begin{verbatim}
@prefix foaf: <http://xmlns.com/foaf/0.1/> .
@prefix rdfs: <http://www.w3.org/2000/01/rdf-schema#> .
<http://example.com/#people> {
  <http://example.com/p#js> a foaf:Person ;
                            foaf:name "John Smith" ;
                            foaf:workplaceHomepage <http://univ.com/> .
  <http://univ.com/> rdfs:label "University" .
}
\end{verbatim}
\label{ex:trig}
\end{example}

\underline{N-Quads} \cite{Carothers2014b} (denoted nq in Table~\ref{tab:rdfserializations}) is another line-based syntax for serializing RDF datasets. 
That plain text format is an RDF syntax similar to N-Triples. N-Quads line statement is a sequence of RDF terms representing the RDF triple (subject, 
predicate and object line in N-Triples) and graph label, which can be an IRI or blank node. The graph label is also a part of a dataset, and this sequence is 
terminated by a dot.
\begin{example}
The example presents a N-Quads serialization that represents RDF triples of Example~\ref{ex:namedgraphex1}. Note that we change some RDF term 
because of legibility.
\begin{verbatim}
<http://univ.com/> <http://...#label> "University" <http://...#pe..>.
<http://example.com/p#js> <http://..#type> <http://../P..> <http://...#pe.>.
<http://example.com/p#js> <http://.../work...>  <.../> <http://...#pe..>.
<http://example.com/p#js> <http://.../name> "John ..." <http://...#pe..>.
\end{verbatim}
\label{ex:nquads}
\end{example}

In Table~\ref{tab:rdfserializations} we present features of the above-mentioned standardized serializations, namely: having W3C Recommendation, human-friendly syntax (partial support means that some fragments may be difficult to read), easy to process, compact form (partial support means that 
there are several different forms and not all are normalized), similarity to Turtle syntax and XML-based syntax and multigraph support. Furthermore, there are a few RDF serializations 
that are not standardized, such as TriX~\cite{carroll2004}, RDF/JSON~\cite{tomaszuk2011,tomaszuk2010}.

\begin{table}[!t]
\caption[RDF serializations]{RDF serializations. This table presents the features of serializations, namely: having W3C Recommendation (yes -- \CheckedBox, no -- \XBox), human-friendly syntax (yes -- \CheckedBox, partial -- \Square, no -- \XBox), easy to process (yes -- \CheckedBox, partial -- \Square, no -- \XBox), compact form (yes -- \CheckedBox, partial -- \Square, no -- \XBox), similarity to Turtle syntax and XML-based syntax (yes -- \CheckedBox, no -- \XBox) and multigraph support (yes -- \CheckedBox, no -- \XBox).}
\label{tab:rdfserializations}
\centering
\begin{tabular}{|l|c|c|c|c|c|c|c|}
\hline
\bf Feature & \bf ttl & \bf nt & \bf tg & \bf nq & \bf jld & \bf rdfa & \bf xml \\
\hline
Standard        & \CheckedBox & \CheckedBox  & \CheckedBox & \CheckedBox  & \CheckedBox     & \CheckedBox & \CheckedBox    \\
Human readable  & \CheckedBox & \Square      & \CheckedBox & \Square      & \CheckedBox     & \CheckedBox & \Square        \\
Efficient       & \XBox       & \XBox        & \XBox       & \XBox        & \Square         & \XBox       & \Square        \\
Normalized      & \XBox       & \CheckedBox  & \XBox       & \CheckedBox  & \Square         & \XBox       & \XBox          \\
Turtle family   & \CheckedBox & \CheckedBox  & \CheckedBox & \CheckedBox  & \XBox           & \XBox       & \XBox          \\
XML family      & \XBox       & \XBox        & \XBox       & \XBox        & \XBox           & \CheckedBox & \CheckedBox    \\
Multiple graphs & \XBox       & \XBox        & \CheckedBox & \CheckedBox  & \CheckedBox     & \XBox       & \XBox          \\
\hline
\end{tabular}
\end{table}

\section{RDF compression}
\label{sec:compression}
RDF compression has been widely addressed recently. However, there is no leading standard for RDF compression.

A recent work \cite{Fernandez2010} points out that RDF datasets are highly compressible because of the RDF graph structure and RDF syntax verbosity. In 
that paper, different compression approaches are analyzed, including:
\begin{enumerate}
 \item direct compression,
 \item adjacency list compression,
 \item an RDF split into the element dictionaries and the statements.
\end{enumerate}
The conclusions of that paper suggest that RDF is highly compressible.

\begin{definition}[RDF compression processor]
\label{def:processor}
An \emph{RDF compression processor} is used by application programs for encoding their RDF data into compressed RDF data and/or to decode 
compressed RDF data to make the data accessible. \qed
\end{definition}

Header-Dictionary-Triples \cite{Fernandez2013} (denoted hdt in Table~\ref{tab:rdfbinserializations}), a binary format that is based on three 
main parts:
\begin{enumerate}
 \item a header, which includes metadata describing the RDF dataset,
 \item a dictionary, which organizes all the identifiers in the graph (it provides a list of the RDF terms such as literals, IRIs and 
blank nodes),
 \item a triples component, which consists of the underlying RDF graph pure structure.
\end{enumerate}
HDT achieves high levels of compression and provides retrieving features to the compressed data. That approach works on the complete 
dataset, with non-negligible processing time. That idea is extended in \cite{Fernandez2014a}. In that thesis, the author proposes 
techniques to compress rich-functional RDF dictionaries and triple indexing. That thesis shows the use of a succinct data configuration 
to browse HDT-encoded datasets. HDT can be used as a backend of Triple Pattern Fragments~\cite{verborgh2014} is natively support fast 
triple-pattern extraction.

In \cite{Fernandez2014b,Alvarez2015} (denoted eri in Table~\ref{tab:rdfbinserializations}), authors exploit a feature of RDF data streams, 
which is the regularity of their data values and structure. They propose a compressed Efficient RDF Interchange format, which can reduce 
the amount of data transmitted when processing RDF streams. ERI considers an RDF stream as a continuous flow of blocks of triples. A 
standard compressor can be used in each channel and impact on its data regularities to produce better compression results.

Another approach is RDF Differential Stream compressor based on Zlib \cite{Fernandez2014c} (denoted rdsz in Table~\ref{tab:rdfbinserializations}), 
which is a proposal for RDF streaming compression. It applies the general-purpose stream compressor Zlib to RDF streams. It uses differential 
encoding to obtain structural similarities. The results of this process are compressed with Zlib to exploit additional redundancies. Furthermore, 
that approach achieves gains in compression at the cost of increasing the processing time.

The interest on RDF compression over streaming data has been indirectly covered by RDF stream processing systems such as Continuous 
Query Evaluation over Linked Streams Cloud \cite{Le2013} (denoted cqels in Table~\ref{tab:rdfbinserializations}) and Ztreamy \cite{Fisteus2014} 
(denoted ztr in Table~\ref{tab:rdfbinserializations}). These papers emphasize the importance of compression for scalable transmission of RDF 
streams over the network. In \cite{Le2013}, authors suggest an approach to deal with this issue by dictionary encoding. In \cite{Fisteus2014}, 
authors discuss a scalable middleware for stream publishing.

Several research areas have emerged around MapReduce and RDF compression, e.g. scalable compression of large RDF datasets \cite{Urbani2013} 
and large RDF data compression and decompression efficiency \cite{Urbani2010} (denoted mr in Table~\ref{tab:rdfbinserializations}). The first paper 
presents an approach based on providing another dictionary-based compression on top of MapReduce \cite{dean2008}. In the second one, authors 
expand \cite{Urbani2010} and achieve linear scalability concerning the number of nodes and input size. Another proposal is presented
in \cite{gimenez2015}, where HDT-MR is introduced. HDT-MR uses MapReduce technique to process huge RDF and build the HDT serialization.

It is worth noting that EXI (Efficient XML Interchange) \cite{kabisch2015} can be used to RDF compression but it can only serialize 
XML~\cite{Schneider2014} i.e. RDF/XML or TriX or JSON~\cite{Peintner2016} i.e. JSON-LD. In Table~\ref{tab:rdfbinserializations} we present 
features of the above-mentioned proposals, namely: having W3C Recommendation, binary syntax, ability to stream, ability to scale 
and support of a software library used for data compression (Zlib).

\begin{table}[!t]
\caption[RDF compression]{RDF compression. This table presents the features of serializations, namely: having W3C Recommendation (yes -- \CheckedBox, no -- \XBox), binary syntax (yes -- \CheckedBox, no -- \XBox), ability to stream (yes -- \CheckedBox, no -- \XBox), ability to scale (yes -- \CheckedBox, no -- \XBox), and support of a software library used for data compression (yes -- \CheckedBox, no -- \XBox)}.
\label{tab:rdfbinserializations}
\centering
\begin{tabular}{|l|c|c|c|c|c|c|}
\hline
\bf Feature   & \bf hdt      & \bf eri     & \bf rdsz    & \bf cqels   & \bf ztr     & \bf mr       \\
\hline
Standard      & \CheckedBox$^\star$ & \XBox       & \XBox       & \XBox       & \XBox       & \XBox        \\ 
Binary format & \CheckedBox         & \CheckedBox & \CheckedBox & \CheckedBox & \CheckedBox & \CheckedBox  \\
Streamable    & \XBox               & \CheckedBox & \CheckedBox & \CheckedBox & \CheckedBox & \CheckedBox  \\
Scalable      & \CheckedBox         & \CheckedBox & \CheckedBox & \CheckedBox & \CheckedBox & \CheckedBox  \\
Zlib          & \XBox               & \XBox       & \CheckedBox & \XBox       & \CheckedBox & \CheckedBox  \\
\hline
\multicolumn{7}{r}{$^\star$ W3C Member Submission}\\
\end{tabular}
\end{table}

\section{Conclusions}
\label{sec:conclusions}
Standards are instrumental in achieving a significant level of interoperability. W3C  recommendations provide people and institutions a basis for mutual understanding. The recommendations that define RDF are used as tools to facilitate various providers to interact with one another. Despite the achievements of current RDF recommendations, they are not sufficient for achieving full end-to-end interoperability. The standards leave several areas vulnerable to variations in interpretation. In this article, we outlined various RDF recommendations, scientific papers that extend and clarify them, and presented a summarised formal description that we hope will clarify some of interpretative differences.

We specifically provided insights on the interpretation of the handling of blank nodes and reification. We presented several interpretative differences, each corresponding to an entailment regime in a standard way. We surveyed various RDF serializations, RDF compression proposals, and RDF mapping approaches to highlight their differences. Finally, we presented a summarized formal definition of RDF 1.1 and emphasized changes between RDF versions 1.0 and 1.1.

We argue that knowledge representation and data integration on the Web faces some of the same challenges we were facing ten years ago, in spite of the significant work being accomplished by both researchers and implementers. We hope that this review contributes to a better understanding of RDF 1.1, and provides the basis for a discussion of interpretative differences. We hope some of these gaps may be able to be fixed in a future version of RDF, such as selection of concise reification, and a formal description of the data model that addresses practical experiences with reification, and blank nodes.

\section*{Acknowledgements}
The authors gratefully acknowledges the members of the RDF~1.1 Working Group who defined RDF version 1.1. We thank Anna Gomolińska for comments that greatly 
improved the manuscript.

\bibliography{rdf11.bib}

\begin{thebibliography}{100}

\bibitem{alkhateeb2008}
Faisal Alkhateeb and S{\'e}bastien Laborie.
\newblock {Towards extending and using {SPARQL} for modular document
  generation}.
\newblock In {\em {Proceedings of the Eighth ACM Symposium on Document
  Engineering}}, {DocEng '08}, pages 164--172, New York, NY, USA, 2008. ACM.

\bibitem{Alvarez2015}
Sandra {\'A}lvarez-Garc\'{\i}a, Nieves~R. Brisaboa, Javier~D. Fern{\'a}ndez,
  Miguel~A. Mart\'{\i}nez-Prieto, and Gonzalo Navarro.
\newblock {Compressed Vertical Partitioning for Efficient RDF Management}.
\newblock {\em Knowledge and Information Systems}, 44(2):439--474, Aug 2015.

\bibitem{amann2002}
Bernd Amann, Catriel Beeri, Irini Fundulaki, and Michel Scholl.
\newblock {Ontology-Based Integration of {XML} Web Resources}.
\newblock In {\em {The Semantic Web --- ISWC 2002}}, {ISWC '02}, pages
  117--131, Berlin, Heidelberg, 2002. Springer Berlin Heidelberg.

\bibitem{antoniou2012}
Grigoris Antoniou, Paul Groth, Frank {Van Harmelen}, and Rinke Hoekstra.
\newblock {\em {A Semantic Web Primer, Third Edition}}.
\newblock MIT press, 2004.

\bibitem{astrova2004}
Irina Astrova.
\newblock {Reverse Engineering of Relational Databases to Ontologies}.
\newblock In {\em {The Semantic Web: Research and Applications}}, pages
  327--341, Berlin, Heidelberg, 2004. Springer Berlin Heidelberg.

\bibitem{astrova2007}
Irina {Astrova}.
\newblock {Rules for Mapping {SQL} Relational Databases to {OWL} Ontologies}.
\newblock In {\em {MTSR}}, pages 415--424, Boston, MA, 2007. Springer US.

\bibitem{Auer2007}
S{\"o}ren Auer, Christian Bizer, Georgi Kobilarov, Jens Lehmann, Richard
  Cyganiak, and Zachary Ives.
\newblock {DBpedia: A Nucleus for a Web of Open Data}.
\newblock In {\em {The Semantic Web}}, pages 722--735, Berlin, Heidelberg,
  2007. Springer Berlin Heidelberg.

\bibitem{auer2009}
S{\"o}ren Auer, Sebastian Dietzold, Jens Lehmann, Sebastian Hellmann, and David
  Aumueller.
\newblock {Triplify: Light-weight Linked Data Publication from Relational
  Databases}.
\newblock In {\em {Proceedings of the 18th International Conference on World
  Wide Web}}, {WWW '09}, pages 621--630, New York, NY, USA, 2009. ACM.

\bibitem{barrasa2004}
Jes{\'u}s Barrasa, {\'O}scar Corcho, and Asunci{\'o}n G{\'o}mez-p{\'e}rez.
\newblock {{R2O}, an Extensible and Semantically based Database-to-Ontology
  Mapping Language}.
\newblock In {\em {in In Proceedings of the 2nd Workshop on Semantic Web and
  Databases(SWDB2004}}, pages 1069--1070. Springer, 2004.

\bibitem{battle2006}
Steve Battle.
\newblock {Gloze: {XML} to {RDF} and back again}.
\newblock In {\em {Jena User Conference}}, 2006.

\bibitem{bedini2011}
Ivan Bedini, Christopher Matheus, Peter~F. Patel-Schneider, Aidan Boran, and
  Benjamin Nguyen.
\newblock {Transforming {XML} Schema to {OWL} Using Patterns}.
\newblock In {\em {Proceedings of the 2011 IEEE Fifth International Conference
  on Semantic Computing}}, {ICSC '11}, pages 102--109, Washington, DC, USA,
  2011. IEEE Computer Society.

\bibitem{berners2003}
Tim Berners-Lee, Dan Connolly, and Sandro Hawke.
\newblock {Semantic Web Tutorial Using N3}.
\newblock In {\em {Twelfth International World Wide Web Conference}}, 2003.

\bibitem{berrueta2008}
Diego Berrueta, Jose~E Labra, and Ivan Herman.
\newblock {{XSLT+SPARQL}: Scripting the Semantic Web with {SPARQL} Embedded
  into {XSLT} Stylesheets}.
\newblock In {\em {4th Workshop on Scripting for the Semantic Web, Tenerife}}.
  Citeseer, 2008.

\bibitem{bikakis2009}
Nikos Bikakis, Nektarios Gioldasis, Chrisa Tsinaraki, and Stavros
  Christodoulakis.
\newblock {Querying {XML} Data with {SPARQL}}.
\newblock In {\em {Database and Expert Systems Applications}}, {DEXA '09},
  pages 372--381, Berlin, Heidelberg, 2009. Springer-Verlag.

\bibitem{Birbeck2010}
Mark Birbeck and Shane McCarron.
\newblock {{CURIE} Syntax 1.0: A syntax for expressing Compact {URIs}}.
\newblock {W3C} working group note, World Wide Web Consortium, 12 2010.

\bibitem{bischof2012}
Stefan Bischof, Stefan Decker, Thomas Krennwallner, Nuno Lopes, and Axel
  Polleres.
\newblock {Mapping between RDF and XML with XSPARQL}.
\newblock {\em Journal on Data Semantics}, 1(3):147--185, Sep 2012.

\bibitem{bischof2011}
Stefan Bischof, Nuno Lopes, and Axel Polleres.
\newblock {Improve Efficiency of Mapping Data Between {XML} and {RDF} with
  {XSPARQL}}.
\newblock In {\em {Proceedings of the 5th International Conference on Web
  Reasoning and Rule Systems}}, {RR'11}, pages 232--237, Berlin, Heidelberg,
  2011. Springer-Verlag.

\bibitem{bizer2004}
Christian Bizer and Andy Seaborne.
\newblock {{D2RQ} - Treating Non-{RDF} Databases as Virtual {RDF} Graphs}.
\newblock In {\em {ISWC2004 (posters)}}, 11 2004.

\bibitem{bohring2005}
Hannes Bohring and S{\"o}ren Auer.
\newblock {Mapping {XML} to {OWL} Ontologies}.
\newblock {\em Leipziger Informatik-Tage}, 72:147--156, 2005.

\bibitem{Boothwell2012}
David Booth.
\newblock {Well Behaved RDF: A Straw-Man Proposal for Taming Blank Nodes},
  2014.

\bibitem{Bray2006}
Tim Bray, Dave Hollander, Andrew Layman, and Richard Tobin.
\newblock {Namespaces in {XML} 1.1 (Second Edition)}.
\newblock {W3C} recommendation, World Wide Web Consortium, 4 2006.

\bibitem{Brickley2014a}
Dan Brickley and Ramanathan Guha.
\newblock {{RDF} Schema {1.1}}.
\newblock {W3C} recommendation, World Wide Web Consortium, 2 2014.

\bibitem{Brickley2014b}
Dan Brickley and Libby Miller.
\newblock {{FOAF} Vocabulary Specification 0.99}.
\newblock Technical report, FOAF Project, 1 2014.

\bibitem{buccella2004}
Agustina Buccella, Miguel~R Penabad, Francisco~J Rodriguez, A~Farina, and
  Alejandra Cechich.
\newblock {From relational databases to {OWL} ontologies}.
\newblock In {\em {Proceedings of the 6th National Russian Research
  Conference}}, 2004.

\bibitem{bumans2010}
Guntars B{\=u}mans and K{\=a}rlis {\v C}er{\=a}ns.
\newblock {{RDB2OWL}: A Practical Approach for Transforming {RDB} Data into
  {RDF/OWL}}.
\newblock In {\em {Proceedings of the 6th International Conference on Semantic
  Systems}}, {I-SEMANTICS '10}, pages 25:1--25:3, New York, NY, USA, 2010. ACM.

\bibitem{buneman2010}
Peter Buneman and Egor Kostylev.
\newblock {Annotation algebras for {RDFS}}.
\newblock In {\em {The Second International Workshop on the role of Semantic
  Web in Provenance Management (SWPM-10), CEUR Workshop Proceedings}}, page~32,
  2010.

\bibitem{byrne2008}
Kate Byrne.
\newblock {Having Triplets -- Holding Cultural Data as RDF}.
\newblock In {\em {Proceedings of the ECDL 2008 Workshop on Information Access
  to Cultural Heritage}}, Aarhus, Denmark, 9 2008.

\bibitem{calvanese2011}
Diego Calvanese, Giuseppe {De Giacomo}, Domenico Lembo, Maurizio Lenzerini,
  Antonella Poggi, Mariano Rodriguez-Muro, Riccardo Rosati, Marco Ruzzi, and
  Domenico~Fabio Savo.
\newblock {The {MASTRO} System for Ontology-based Data Access}.
\newblock {\em Semantic Web}, 2(1):43--53, 1 2011.

\bibitem{Carothers2014b}
Gavin Carothers.
\newblock {{RDF} 1.1 N-Quads}.
\newblock {W3C} recommendation, World Wide Web Consortium, 2 2014.

\bibitem{Carothers2014a}
Gavin Carothers and Andy Seaborne.
\newblock {{RDF} 1.1 N-Triples}.
\newblock {W3C} recommendation, World Wide Web Consortium, 2 2014.

\bibitem{carroll2002}
Jeremy~J. Carroll.
\newblock {Matching {RDF} Graphs}.
\newblock In {\em {The Semantic Web --- ISWC 2002}}, pages 5--15, Berlin,
  Heidelberg, 2002. Springer Berlin Heidelberg.

\bibitem{carroll2005}
Jeremy~J Carroll, Christian Bizer, Pat Hayes, and Patrick Stickler.
\newblock {Named Graphs, Provenance and Trust}.
\newblock In {\em {Proceedings of the 14th International Conference on World
  Wide Web}}, {WWW '05}, pages 613--622, New York, NY, USA, 2005. ACM.

\bibitem{carroll2004}
Jeremy~J Carroll and Patrick Stickler.
\newblock {{RDF} triples in {XML}}.
\newblock In {\em {Proceedings of the 13th International World Wide Web
  Conference on Alternate Track Papers \& Posters}}, {WWW Alt. '04}, pages
  412--413, New York, NY, USA, 2004. ACM.

\bibitem{cerbah2008a}
Farid Cerbah.
\newblock {Learning Highly Structured Semantic Repositories from Relational
  Databases: The RDBToOnto Tool}.
\newblock In {\em {The Semantic Web: Research and Applications}}, {ESWC'08},
  pages 777--781, Berlin, Heidelberg, 2008. Springer-Verlag.

\bibitem{cerbah2008b}
{Farid} {Cerbah}.
\newblock {Mining the Content of Relational Databases to Learn Ontologies with
  Deeper Taxonomies}.
\newblock In {\em {Web Intelligence}}, volume~1, pages 553--557. IEEE Computer
  Society, Dec 2008.

\bibitem{ceri1991}
Stefano Ceri, Letizia Tanca, and Roberto~V. Zicari.
\newblock {Supporting Interoperability Between New Database Languages}.
\newblock In {\em {Proc. of the 5th Annual European Computer Conference
  (CompEuro)}}, pages 273--281, 5 1991.

\bibitem{champin2001}
Pierre-Antoine Champin.
\newblock {RDF tutorial}, 2001.

\bibitem{Chen2012}
Lei Chen, Haifei Zhang, Ying Chen, and Wenping Guo.
\newblock {Blank Nodes in RDF}.
\newblock {\em Journal of Software}, 7(9):1993--1999, 2012.

\bibitem{chinnici2007}
Roberto Chinnici, Jean-Jacques Moreau, Arthur Ryman, and Sanjiva Weerawarana.
\newblock {Web Services Description Language ({WSDL}) Version 2.0 Part 1: Core
  Language}.
\newblock {W3C} recommendation, World Wide Web Consortium, 6 2007.

\bibitem{connolly2007}
Dan Connolly.
\newblock {Gleaning Resource Descriptions from Dialects of Languages
  ({GRDDL})}.
\newblock {W3C} recommendation, World Wide Web Consortium, 9 2007.

\bibitem{corby2015a}
Olivier Corby and Catherine Faron-Zucker.
\newblock {A Transformation Language for {RDF} based on {SPARQL}}.
\newblock In {\em {Web Information Systems and Technologies}}, pages 318--340,
  Cham, 2016. Springer International Publishing.

\bibitem{corby2015b}
Olivier Corby, Catherine Faron-Zucker, and Fabien Gandon.
\newblock {A Generic {RDF} Transformation Software and Its Application to an
  Online Translation Service for Common Languages of Linked Data}.
\newblock In {\em {International Semantic Web Conference (2)}}, volume 9367 of
  {\em {Lecture Notes in Computer Science}}, pages 150--165. Springer, 2015.

\bibitem{cruz2008}
Christophe Cruz and Christophe Nicolle.
\newblock {Ontology Enrichment and Automatic Population From {XML} Data}.
\newblock In {\em {Proceedings of the 4th International VLDB Workshop on
  Ontology-based Techniques for DataBases in Information Systems and Knowledge
  Systems, ODBIS 2008}}, pages 17--20, Auckland, New Zealand, 8 2008.

\bibitem{Crystal1997}
David Crystal.
\newblock {\em {The Cambridge Encyclopedia of Language (3rd edition)}}.
\newblock Cambridge University Press, 11 2018.

\bibitem{cure2014}
Olivier Cur{\'e} and Guillaume Blin.
\newblock {\em {RDF database systems: triples storage and {SPARQL} query
  processing}}.
\newblock Morgan Kaufmann, 2014.

\bibitem{curino2009}
Carlo Curino, Giorgio Orsi, Emanuele Panigati, and Letizia Tanca.
\newblock {Accessing and Documenting Relational Databases Through {OWL}
  Ontologies}.
\newblock In {\em {Proceedings of the 8th International Conference on Flexible
  Query Answering Systems}}, {FQAS '09}, pages 431--442, Berlin, Heidelberg,
  2009. Springer-Verlag.

\bibitem{Cyganiak2014}
Richard Cyganiak, Markus Lanthaler, and David Wood.
\newblock {{RDF} 1.1 Concepts and Abstract Syntax}.
\newblock {W3C} recommendation, World Wide Web Consortium, 2 2014.

\bibitem{das2012}
Souripriya Das, Richard Cyganiak, and Seema Sundara.
\newblock {{R2RML}: {RDB} to {RDF} Mapping Language}.
\newblock {W3C} recommendation, World Wide Web Consortium, 9 2012.

\bibitem{date1975}
CJ~Data.
\newblock {\em {An Introduction to Database Systems}}.
\newblock Addison-Wesley publ., 1975.

\bibitem{bruijn2005}
Jos~J. de~Bruijn, Enrico Franconi, and Sergio Tessaris.
\newblock {Logical Reconstruction of {RDF} and Ontology Languages}.
\newblock In {\em {Principles and Practice of Semantic Web Reasoning}}, pages
  65--71, Berlin, Heidelberg, 2005. Springer Berlin Heidelberg.

\bibitem{delaborda2005b}
Cristian~P{\'e}rez {De Laborda} and Stefan Conrad.
\newblock {Querying Relational Databases with {RDQL}}.
\newblock In {\em {Berliner XML Tage}}, pages 161--172. Citeseer, 2005.

\bibitem{delaborda2005a}
Cristian~P{\'e}rez de~Laborda and Stefan Conrad.
\newblock {Relational.OWL: A Data and Schema Representation Format Based on
  OWL}.
\newblock In {\em {Proceedings of the 2Nd Asia-Pacific Conference on Conceptual
  Modelling - Volume 43}}, {APCCM '05}, pages 89--96, Darlinghurst, Australia,
  Australia, 2005. Australian Computer Society, Inc.

\bibitem{dean2008}
Jeffrey Dean and Sanjay Ghemawat.
\newblock {MapReduce: Simplified Data Processing on Large Clusters}.
\newblock {\em Communications of the ACM}, 51(1):107--113, 2008.

\bibitem{decker2000}
Stefan Decker, Prasenjit Mitra, and Sergey Melnik.
\newblock {Framework for the Semantic Web: an {RDF} Tutorial}.
\newblock {\em IEEE Internet Computing}, 4(6):68--73, Nov 2000.

\bibitem{deursen2008}
Davy~Van Deursen, Chris Poppe, G{\"a}etan Martens, Erik Mannens, and Rik Van~de
  Walle.
\newblock {{XML} to {RDF} Conversion: A Generic Approach}.
\newblock In {\em {Proceedings of the 2008 International Conference on
  Automated Solutions for Cross Media Content and Multi-channel Distribution}},
  {AXMEDIS '08}, pages 138--144, Washington, DC, USA, 11 2008. IEEE Computer
  Society.

\bibitem{dimou2014}
Anastasia Dimou, Miel {Vander Sande}, Pieter Colpaert, Ruben Verborgh, Erik
  Mannens, and Rik {Van de Walle}.
\newblock {{RML}: A Generic Language for Integrated {RDF} Mappings of
  Heterogeneous Data}.
\newblock In {\em {Proceedings of the 7th Workshop on Linked Data on the Web}},
  volume 1184 of {\em {CEUR Workshop Proceedings}}, 4 2014.

\bibitem{droop2009}
Matthias Droop, Markus Flarer, Jinghua Groppe, Sven Groppe, Volker Linnemann,
  Jakob Pinggera, Florian Santner, Michael Schier, Felix Sch{\"o}pf, Hannes
  Staffler, et~al.
\newblock {Bringing the {XML} and Semantic Web Worlds Closer: Transforming
  {XML} into {RDF} and embedding {XPath} into {SPARQL}}.
\newblock In {\em {Enterprise Information Systems}}, pages 31--45. Springer
  Berlin Heidelberg, Berlin, Heidelberg, 2009.

\bibitem{erling2007}
Orri Erling and Ivan Mikhailov.
\newblock {{RDF} Support in the Virtuoso {DBMS}}.
\newblock In {\em {Networked Knowledge - Networked Media: Integrating Knowledge
  Management, New Media Technologies and Semantic Systems}}, pages 7--24,
  Berlin, Heidelberg, 2009. Springer Berlin Heidelberg.

\bibitem{erling2009}
Orri Erling and Ivan Mikhailov.
\newblock {{RDF} Support in the Virtuoso {DBMS}}.
\newblock In {\em {Networked Knowledge-Networked Media}}, pages 7--24.
  Springer, 2009.

\bibitem{farrell2007}
Joel Farrell and Holger Lausen.
\newblock {Semantic Annotations for {WSDL} and {XML} Schema}.
\newblock {W3C} recommendation, World Wide Web Consortium, 8 2007.

\bibitem{feigenbaum2007}
Lee Feigenbaum, Ivan Herman, Tonya Hongsermeier, Eric Neumann, and Susie
  Stephens.
\newblock {The Semantic Web in Action}.
\newblock {\em Scientific American}, 297(6):90--97, 2007.

\bibitem{ferdinand2004}
Matthias Ferdinand, Christian Zirpins, and David Trastour.
\newblock {Lifting {XML} schema to {OWL}}.
\newblock In {\em {Web Engineering}}, pages 354--358. Springer Berlin
  Heidelberg, Berlin, Heidelberg, 2004.

\bibitem{Fernandez2014a}
Javier~D. Fern{\'a}ndez.
\newblock {Binary {RDF} for Scalable Publishing, Exchanging and Consumption in
  the Web of Data}.
\newblock In {\em {Proceedings of the 21st International Conference on World
  Wide Web}}, {WWW '12 Companion}, pages 133--138, New York, NY, USA, 2012.
  ACM.

\bibitem{Fernandez2010}
Javier~D. Fern{\'a}ndez, Claudio Gutierrez, and Miguel~A. Mart\'{\i}nez-Prieto.
\newblock {RDF Compression: Basic Approaches}.
\newblock In {\em {Proceedings of the 19th International Conference on World
  Wide Web}}, {WWW '10}, pages 1091--1092, New York, NY, USA, 2010. ACM.

\bibitem{Fernandez2014b}
Javier~D. Fern{\'a}ndez, Alejandro Llaves, and Oscar Corcho.
\newblock {Efficient RDF Interchange (ERI) Format for RDF Data Streams}.
\newblock In {\em {The Semantic Web -- ISWC 2014}}, pages 244--259, Cham, 2014.
  Springer International Publishing.

\bibitem{Fernandez2013}
Javier~D Fern{\'a}ndez, Miguel~A Mart\'{\i}nez-Prieto, Claudio Guti{\'e}rrez,
  Axel Polleres, and Mario Arias.
\newblock {Binary {RDF} Representation for Publication and Exchange ({HDT})}.
\newblock {\em Journal of Web Semantics}, 19:22--41, 2013.

\bibitem{Fernandez2014c}
Norberto Fern{\'a}ndez, Jes{\'u}s Arias, Luis S{\'a}nchez, Damaris
  Fuentes-Lorenzo, and {\'O}scar Corcho.
\newblock {{RDSZ}: An approach for lossless {RDF} stream compression}.
\newblock In {\em {The Semantic Web: Trends and Challenges}}, pages 52--67,
  Cham, 2014. Springer International Publishing.

\bibitem{Fisteus2014}
Jesus~Arias Fisteus, Norberto~Fern{\'a}ndez Garcia, Luis~S{\'a}nchez Fernandez,
  and Damaris Fuentes-Lorenzo.
\newblock {Ztreamy: A middleware for publishing semantic streams on the Web}.
\newblock {\em Journal of Web Semantics}, 25:16--23, 2014.

\bibitem{franconi2005}
Enrico Franconi, Jos de~Bruijn, and Sergio Tessaris.
\newblock {Logical Reconstruction of Normative {RDF}}.
\newblock In {\em {OWLED}}, volume 188 of {\em {CEUR Workshop Proceedings}}.
  CEUR-WS.org, 2005.

\bibitem{Gandon2014}
Fabien Gandon and Guus Schreiber.
\newblock {{RDF} 1.1 {XML} Syntax}.
\newblock {W3C} recommendation, World Wide Web Consortium, 2 2014.

\bibitem{garcia2005}
Roberto Garcia and Oscar Celma.
\newblock {Semantic Integration and Retrieval of Multimedia Metadata}.
\newblock In {\em {5th International Workshop on Knowledge Markup and Semantic
  Annotation}}, pages 69--80, 2005.

\bibitem{ghawi2007}
Raji Ghawi and Nadine Cullot.
\newblock {Database-to-Ontology Mapping Generation for Semantic
  Interoperability}.
\newblock In {\em {Third International Workshop on Database Interoperability
  (InterDB 2007)}}, 09 2007.

\bibitem{ghawi2009}
Raji Ghawi and Nadine Cullot.
\newblock {Building Ontologies from {XML} Data Sources}.
\newblock In {\em {Proceedings of the 2009 20th International Workshop on
  Database and Expert Systems Application}}, {DEXA '09}, pages 480--484,
  Washington, DC, USA, Aug 2009. IEEE Computer Society.

\bibitem{gimenez2015}
Jos{\'e}~M Gim{\'e}nez-Garc\'{\i}a, Javier~D Fern{\'a}ndez, and Miguel~A
  Mart\'{\i}nez-Prieto.
\newblock {{HDT-MR}: A scalable solution for RDF compression with HDT and
  MapReduce}.
\newblock In {\em {The Semantic Web. Latest Advances and New Domains}}, pages
  253--268, Cham, 2015. Springer International Publishing.

\bibitem{grau2004}
Bernardo~Cuenca Grau.
\newblock {A Possible Simplification of the Semantic Web Architecture}.
\newblock In {\em {Proceedings of the 13th International Conference on World
  Wide Web}}, {WWW '04}, pages 704--713, New York, NY, USA, 2004. ACM.

\bibitem{groth2010}
Paul Groth, Andrew Gibson, and Jan Velterop.
\newblock {The anatomy of a Nano-publication}.
\newblock {\em Information Services and Use}, 30(1-2):51--56, 2010.

\bibitem{gutierrez2005}
Claudio Gutierrez, Carlos Hurtado, and Alejandro Vaisman.
\newblock {Temporal RDF}.
\newblock In {\em {The Semantic Web: Research and Applications}}, pages
  93--107. Springer Berlin Heidelberg, Berlin, Heidelberg, 2005.

\bibitem{gutierrez2004}
Claudio Gutierrez, Carlos~A. Hurtado, and Alberto~O. Mendelzon.
\newblock {Foundations of Semantic Web Databases}.
\newblock In {\em {ACM Symposium on Principles of Database Systems (PODS)}},
  pages 95--106, 2004.

\bibitem{Gutierrez2011}
Claudio Gutierrez, Carlos~A. Hurtado, Alberto~O. Mendelzon, and P{\'e}rez
  Jorge.
\newblock {Foundations of Semantic Web Databases}.
\newblock {\em Journal of Computer and System Sciences}, 77(3):520--541, 2011.
\newblock Database Theory.

\bibitem{harris2013}
Steven Harris and Andy Seaborne.
\newblock {{SPARQL} 1.1 Query Language}.
\newblock {W3C} recommendation, World Wide Web Consortium, 3 2013.

\bibitem{harth2014}
Andreas Harth, Katja Hose, and Ralf Schenkel.
\newblock {\em {Linked Data Management}}.
\newblock CRC Press, 2014.

\bibitem{hartig2014}
Olaf Hartig and Bryan Thompson.
\newblock {Foundations of an Alternative Approach to Reification in {RDF}}.
\newblock {\em CoRR}, abs/1406.3399, 2014.

\bibitem{Hayes2004}
Patrick Hayes.
\newblock {{RDF} 1.1 Semantics}.
\newblock {W3C} recommendation, World Wide Web Consortium, 2 2014.

\bibitem{hert2010}
Matthias Hert, Gerald Reif, and Harald~C. Gall.
\newblock {Updating Relational Data via {SPARQL/Update}}.
\newblock In {\em {Proceedings of the 2010 EDBT/ICDT Workshops}}, {EDBT '10},
  pages 24:1--24:8, New York, NY, USA, 2010. ACM.

\bibitem{Hitzler2011}
Pascal Hitzler, Markus Krotzsch, and Sebastian Rudolph.
\newblock {\em {Foundations of Semantic Web Technologies}}.
\newblock Chapman \& Hall/CRC, 1st edition, 2011.

\bibitem{hogan2015}
Aidan Hogan.
\newblock {Skolemising Blank Nodes While Preserving Isomorphism}.
\newblock In {\em {Proceedings of the 24th International Conference on World
  Wide Web}}, {WWW '15}, pages 430--440, Republic and Canton of Geneva,
  Switzerland, 2015. International World Wide Web Conferences Steering
  Committee.

\bibitem{Hogan2015b}
Aidan Hogan.
\newblock {Skolemising Blank Nodes while Preserving Isomorphism}.
\newblock In {\em {Proceedings of the 24th International Conference on World
  Wide Web}}, {WWW '15}, pages 430--440, Republic and Canton of Geneva,
  Switzerland, 2015. International World Wide Web Conferences Steering
  Committee.

\bibitem{Hogan2014}
Aidan Hogan, Marcelo Arenas, Alejandro Mallea, and Axel Polleres.
\newblock {Everything You Always Wanted to Know About Blank Nodes}.
\newblock {\em Journal of Web Semantics}, 27-28:42--69, 2014.
\newblock Semantic Web Challenge 2013.

\bibitem{patel-schneider2002}
Ian Horrocks, Peter~F. Patel-Schneider, and Frank~Van Harmelen.
\newblock {Reviewing the Design of {DAML+OIL}: An Ontology Language for the
  Semantic Web}.
\newblock In {\em {Eighteenth National Conference on Artificial Intelligence}},
  pages 792--797, Menlo Park, CA, USA, 2002. American Association for
  Artificial Intelligence.

\bibitem{hu2007}
Wei Hu and Yuzhong Qu.
\newblock {Discovering Simple Mappings Between Relational Database Schemas and
  Ontologies}.
\newblock In {\em {The Semantic Web}}, pages 225--238, Berlin, Heidelberg,
  2007. Springer Berlin Heidelberg.

\bibitem{kabisch2015}
Sebastian K{\"a}bisch, Daniel Peintner, and Darko Anicic.
\newblock {Standardized and Efficient {RDF} Encoding for Constrained Embedded
  Networks}.
\newblock In {\em {The Semantic Web. Latest Advances and New Domains}}, pages
  437--452, Cham, 2015. Springer International Publishing.

\bibitem{kay2007}
Michael Kay.
\newblock {{XSL} Transformations ({XSLT}) Version 3.0}.
\newblock {W3C} recommendation, World Wide Web Consortium, 6 2017.

\bibitem{kifer1989}
Michael Kifer and Georg Lausen.
\newblock {F-logic: a Higher-order Language for Reasoning About Objects,
  Inheritance, and Scheme}.
\newblock {\em SIGMOD Rec.}, 18(2):134--146, 6 1989.

\bibitem{klein2002}
Michel C.~A. Klein.
\newblock {Interpreting {XML} Documents via an {RDF} Schema Ontology}.
\newblock In {\em {Proceedings of the 13th International Workshop on Database
  and Expert Systems Applications}}, {DEXA '02}, pages 889--894, Washington,
  DC, USA, Sep 2002. IEEE Computer Society.

\bibitem{koffina2006}
Ioanna Koffina, Giorgos Serfiotis, Vassilis Christophides, and Val Tannen.
\newblock {Mediating {RDF/S} Queries to Relational and {XML} Sources}.
\newblock {\em International Journal on Semantic Web and Information Systems},
  2(4):68, 2006.

\bibitem{koubarakis2010}
Manolis Koubarakis and Kostis Kyzirakos.
\newblock {Modeling and Querying Metadata in the Semantic Sensor Web: The Model
  stRDF and the Query Language {stSPARQL}}.
\newblock In {\em {ESWC (1)}}, volume 6088 of {\em {Lecture Notes in Computer
  Science}}, pages 425--439. Springer, 2010.

\bibitem{Lassila1999}
Ora Lassila and Ralph~R. Swick.
\newblock {Resource Description Framework ({RDF}) Model and Syntax
  Specification}.
\newblock {W3C} recommendation, World Wide Web Consortium, 2 1999.

\bibitem{Le2013}
Danh Le-Phuoc, Hoan Nguyen~Mau Quoc, Chan {Le Van}, and Manfred Hauswirth.
\newblock {Elastic and Scalable Processing of Linked Stream Data in the Cloud}.
\newblock In {\em {The Semantic Web -- ISWC 2013}}, pages 280--297, Berlin,
  Heidelberg, 2013. Springer Berlin Heidelberg.

\bibitem{lehti2004}
Patrick Lehti and Peter Fankhauser.
\newblock {{XML} Data Integration with {OWL}: Experiences and Challenges}.
\newblock In {\em {Applications and the Internet, 2004. Proceedings. 2004
  International Symposium on}}, pages 160--167. IEEE, Jan 2004.

\bibitem{lenzerini2002}
Maurizio Lenzerini.
\newblock {Data Integration: A Theoretical Perspective}.
\newblock In {\em {Proceedings of the Twenty-first ACM SIGMOD-SIGACT-SIGART
  Symposium on Principles of Database Systems}}, {PODS '02}, pages 233--246,
  New York, NY, USA, 2002. ACM.

\bibitem{li2005}
Man Li, Xiaoyong Du, and Shan Wang.
\newblock {A Semi-automatic Ontology Acquisition Method for the Semantic Web}.
\newblock In {\em {Advances in Web-Age Information Management}}, {WAIM'05},
  pages 209--220, Berlin, Heidelberg, 2005. Springer-Verlag.

\bibitem{lopes2011}
Nuno Lopes, Stefan Bischof, Stefan Decker, and Axel Polleres.
\newblock {On the Semantics of Heterogeneous Querying of Relational, {XML} and
  {RDF} Data with {XSPARQL}}.
\newblock In {\em {Proceedings of the 15th Portuguese Conference on Artificial
  Intelligence (EPIA 2011), Lisbon, Portugal}}, 2011.

\bibitem{Mallea2011}
Alejandro Mallea, Marcelo Arenas, Aidan Hogan, and Axel Polleres.
\newblock {On Blank Nodes}.
\newblock In {\em {The Semantic Web -- ISWC 2011}}, pages 421--437, Berlin,
  Heidelberg, 2011. Springer Berlin Heidelberg.

\bibitem{Manola2004}
Frank Manola and Eric Miller.
\newblock {RDF Primer}.
\newblock {W3C} recommendation, World Wide Web Consortium, 2 2004.

\bibitem{Marin2004}
Draltan Marin.
\newblock {A formalization of RDF}.
\newblock Master's thesis, {\'E}cole Polytechnique, 8 2004.

\bibitem{McCarron2015}
Shane McCarron, Ben Adida, Mark Birbeck, and Ivan Herman.
\newblock {{RDFa} Core {1.1} - Third Edition}.
\newblock {W3C} recommendation, World Wide Web Consortium, 8 2015.

\bibitem{Munoz2007}
Sergio Mu{\~n}oz, Jorge P{\'e}rez, and Claudio Gutierrez.
\newblock {Minimal Deductive Systems for RDF}.
\newblock In {\em {The Semantic Web: Research and Applications}}, pages 53--67,
  Berlin, Heidelberg, 2007. Springer Berlin Heidelberg.

\bibitem{munoz2009}
Sergio Munoz, Jorge Perez, and Claudio Gutierrez.
\newblock {Simple and Efficient Minimal {RDFS}}.
\newblock {\em Web Semant.}, 7(3):220--234, 9 2009.

\bibitem{nguyen2014}
Vinh Nguyen, Olivier Bodenreider, and Amit Sheth.
\newblock {Don't Like {RDF} Reification?: Making Statements About Statements
  Using Singleton Property}.
\newblock In {\em {Proceedings of the 23rd International Conference on World
  Wide Web}}, {WWW '14}, pages 759--770, New York, NY, USA, 2014. ACM.

\bibitem{Nottingham2010}
Mark Nottingham and Eran Hammer-Lahav.
\newblock {Defining Well-Known Uniform Resource Identifiers ({URI}s)}.
\newblock {RFC} 5785, Internet Engineering Task Force, 4 2010.

\bibitem{Noy2006}
Natasha Noy and Alan Rector.
\newblock {Defining {N}-ary Relations on the Semantic Web}.
\newblock {W3C} working group note, World Wide Web Consortium, 4 2006.

\bibitem{nyulas2007}
Csongor Nyulas, Martin OConnor, and Samson Tu.
\newblock {DataMaster--a plug-in for importing schemas and data from relational
  databases into Protege}.
\newblock In {\em {Proceedings of the 10th International Protege Conference}}.
  Citeseer, 2007.

\bibitem{oconnor2011}
Martin~J O'Connor and Amar Das.
\newblock {Acquiring OWL Ontologies from XML Documents}.
\newblock In {\em {Proceedings of the Sixth International Conference on
  Knowledge Capture}}, {K-CAP '11}, pages 17--24, New York, NY, USA, 2011. ACM.

\bibitem{Pan2003}
Jeff~Z. Pan and Ian Horrocks.
\newblock {{RDFS(FA)} and {RDF MT}: Two semantics for {RDFS}}.
\newblock In {\em {The Semantic Web - ISWC 2003}}, pages 30--46, Berlin,
  Heidelberg, 2003. Springer Berlin Heidelberg.

\bibitem{Patel-Schneider2014}
Peter Patel-Schneider and Patrick Hayes.
\newblock {{RDF} 1.1 Semantics}.
\newblock {W3C} recommendation, World Wide Web Consortium, 2 2014.

\bibitem{Peintner2016}
Daniel Peintner and Don Brutzman.
\newblock {{EXI} for {JSON} ({EXI4JSON})}.
\newblock {W3C} working group note, World Wide Web Consortium, 4 2018.

\bibitem{peroni2015}
Silvio Peroni and Fabio Vitali.
\newblock {{RSLT}: {RDF} Stylesheet Language Transformations}.
\newblock In {\em {ESWC Developers Workshop}}, volume 1361 of {\em {CEUR
  Workshop Proceedings}}, pages 7--13. CEUR-WS.org, 2015.

\bibitem{Pichler2008}
Reinhard Pichler, Axel Polleres, Fang Wei, and Stefan Woltran.
\newblock {{dRDF}: Entailment for Domain-restricted {RDF}}.
\newblock In {\em {The Semantic Web: Research and Applications}}, pages
  200--214, Berlin, Heidelberg, 2008. Springer Berlin Heidelberg.

\bibitem{polflietI2010}
Simeon Polfliet and Ryutaro Ichise.
\newblock {Automated Mapping Generation for Converting Databases into Linked
  Data}.
\newblock In {\em {ISWC Posters\&Demos}}, volume 658 of {\em {CEUR Workshop
  Proceedings}}. CEUR-WS.org, 2010.

\bibitem{priyatna2015}
Freddy Priyatna, Ra{\'u}l Alonso-Calvo, Sergio Paraiso-Medina, Gueton
  Padron-Sanchez, and Oscar Corcho.
\newblock {{R2RML}-based Access and Querying to Relational Clinical Data with
  Morph-RDB}.
\newblock In {\em {SWAT4LS}}, pages 142--151, 2015.

\bibitem{Prud'hommeaux2014}
Eric Prud'hommeaux and Gavin Carothers.
\newblock {{RDF} 1.1 Turtle}.
\newblock {W3C} recommendation, World Wide Web Consortium, 2 2014.

\bibitem{quan2005}
Dennis Quan and David~R. Karger.
\newblock {Xenon: An {RDF} Stylesheet Ontology}.
\newblock {\em Procs of WWW}, page~40, 2005.

\bibitem{Raimond2014}
Yves Raimond and Guus Schreiber.
\newblock {{RDF} 1.1 Primer}.
\newblock {W3C} note, World Wide Web Consortium, 2 2014.

\bibitem{reif2004}
Gerald Reif, Mehdi Jazayeri, and Harald Gall.
\newblock {Towards Semantic Web Engineering: {WEESA}-Mapping {XML} Schema to
  Ontologies}.
\newblock In {\em {WWW Workshop on Application Design, Development and
  Implementation Issues in the Semantic Web}}, 2004.

\bibitem{robie2014}
Jonathan Robie and Michael Dyck.
\newblock {{XQuery} 3.1: An XML Query Language}.
\newblock {W3C} recommendation, World Wide Web Consortium, 3 2017.

\bibitem{rodrigues2008}
Toni Rodrigues, Pedro Rosa, and Jorge Cardoso.
\newblock {Moving from Syntactic to Semantic Organizations Using {JXML2OWL}}.
\newblock {\em Comput. Ind.}, 59(8):808--819, 10 2008.

\bibitem{sahoo2008}
Satya~S. Sahoo, Olivier Bodenreider, Joni~L. Rutter, Karen~J. Skinner, and
  Amit~P. Sheth.
\newblock {An Ontology-driven Semantic Mashup of Gene and Biological Pathway
  Information: Application to the Domain of Nicotine Dependence}.
\newblock {\em J. of Biomedical Informatics}, 41(5):752--765, 10 2008.

\bibitem{salas2010}
Percy~E. Salas, Karin~K. Breitman, Jos{\'e} {Viterbo F.}, and Marco~A.
  Casanova.
\newblock {Interoperability by Design Using the StdTrip Tool: An a Priori
  Approach}.
\newblock In {\em {Proceedings of the 6th International Conference on Semantic
  Systems}}, {I-SEMANTICS '10}, pages 43:1--43:3, New York, NY, USA, 2010. ACM.

\bibitem{Schneider2014}
John Schneider, Takuki Kamiya, Daniel Peintner, and Rumen Kyusakov.
\newblock {Efficient {XML} Interchange ({EXI}) Format 1.0 (Second Edition)}.
\newblock {W3C} recommendation, World Wide Web Consortium, 2 2014.

\bibitem{schwitter2004}
Rolf Schwitter, Marc Tilbrook, et~al.
\newblock {Controlled Natural Language Meets the Semantic Web}.
\newblock In {\em {Proceedings of the Australasian Language Technology
  Workshop}}, volume~2, pages 55--62. Australian Speech Science and Technology
  Association, 2004.

\bibitem{Seaborne2014}
Andy Seaborne and Gavin Carothers.
\newblock {{RDF} 1.1 TriG}.
\newblock {W3C} recommendation, World Wide Web Consortium, 2 2014.

\bibitem{seaborne2007}
Andy Seaborne, Damian Steer, and Stuart Williams.
\newblock {SQL-RDF}.
\newblock http://www.w3.org/2007/03/RdfRDB/papers/seaborne.html, October 2007.

\bibitem{sequeda2015}
Juan~F Sequeda and Daniel~P Miranker.
\newblock {Ultrawrap Mapper: A Semi-Automatic Relational Database to {RDF}
  ({RDB2RDF}) Mapping Tool}.
\newblock In {\em {International Semantic Web Conference (Posters \& Demos)}},
  2015.

\bibitem{sequeda2009}
Juan~F Sequeda, Syed~H Tirmizi, Oscar Corcho, and Daniel~P Miranker.
\newblock {Direct mapping {SQL} Databases to the Semantic Web: A Survey}.
\newblock {\em Univeristy of Texas, Department of Computer Sciecnces Technical
  Report TR-09-04}, 2009.

\bibitem{shapkin2015}
Pavel Shapkin and Leonid Shumsky.
\newblock {A Language for Transforming the {RDF} Data on the Basis of
  Ontologies}.
\newblock In {\em {Proceedings of the 11th International Conference on Web
  Information Systems and Technologies}}, pages 504--511, 2015.

\bibitem{shen2006}
Guohua Shen, Zhiqiu Huang, Xiaodong Zhu, and Xiaofei Zhao.
\newblock {Research on the Rules of Mapping from Relational Model to {OWL}}.
\newblock In {\em {OWLED}}, volume 216 of {\em {CEUR Workshop Proceedings}}.
  CEUR-WS.org, 2006.

\bibitem{snelson2014}
John Snelson, Don Chamberlin, Michael Dyck, and Jonathan Robie.
\newblock {{XML} Path Language ({XPath}) 3.1}.
\newblock {W3C} recommendation, World Wide Web Consortium, 3 2017.

\bibitem{Sowa1999}
John~F. Sowa.
\newblock {\em {Knowledge Representation: Logical, Philosophical, and
  Computational Foundations}}.
\newblock Brooks/Cole Publishing Co., Pacific Grove, CA, USA, 2000.

\bibitem{spanos2012}
Dimitrios-Emmanuel Spanos, Periklis Stavrou, and Nikolas Mitrou.
\newblock {Bringing Relational Databases into the Semantic Web: A Survey}.
\newblock {\em Semantic Web}, 3(2):169--209, 4 2012.

\bibitem{Sperberg-McQueen2013}
Michael Sperberg-McQueen, Henry Thompson, David Peterson, Ashok Malhotra,
  Paul~V. Biron, and Sandy Gao.
\newblock {{W3C} {XML} Schema Definition Language ({XSD}) {1.1} Part 2:
  Datatypes}.
\newblock {W3C} recommendation, World Wide Web Consortium, 4 2012.

\bibitem{Sporny2014}
Manu Sporny, Markus Lanthaler, and Gregg Kellogg.
\newblock {{JSON}-LD 1.0}.
\newblock {W3C} recommendation, World Wide Web Consortium, 1 2014.

\bibitem{stavrakantonakis2010}
Ioannis Stavrakantonakis, Chrisa Tsinaraki, Nikos Bikakis, Nektarios Gioldasis,
  and Stavros Christodoulakis.
\newblock {{SPARQL2XQuery 2.0}: Supporting Semantic-based queries over {XML}
  data}.
\newblock In {\em {SMAP}}, pages 76--84. IEEE, 2010.

\bibitem{stojanovic2002}
Ljiljana Stojanovic, Nenad Stojanovic, and Raphael Volz.
\newblock {Migrating Data-intensive Web Sites into the Semantic Web}.
\newblock In {\em {Proceedings of the 2002 ACM Symposium on Applied
  Computing}}, {SAC '02}, pages 1100--1107, New York, NY, USA, 2002. ACM.

\bibitem{straccia2009}
Umberto Straccia.
\newblock {A minimal deductive system for general fuzzy {RDF}}.
\newblock In {\em {Web Reasoning and Rule Systems}}, pages 166--181. Springer
  Berlin Heidelberg, Berlin, Heidelberg, 2009.

\bibitem{Tandy2015}
Jeremy Tandy, Ivan Herman, and Gregg Kellogg.
\newblock {Generating {RDF} from Tabular Data on the Web}.
\newblock {W3C} recommendation, World Wide Web Consortium, 12 2015.

\bibitem{Terhorst2005}
Herman~J. {ter Horst}.
\newblock {Completeness, Decidability and Complexity of Entailment for {RDF
  Schema} and a Semantic Extension Involving the {OWL} Vocabulary}.
\newblock {\em Journal of Web Semantics}, 3(2):79--115, 2005.
\newblock Selcted Papers from the International Semantic Web Conference, 2004.

\bibitem{thakkar2019}
Harsh Thakkar, Renzo Angles, Dominik Tomaszuk, and Jens Lehmann.
\newblock {Direct Mappings between RDF and Property Graph Databases}.
\newblock {\em arXiv}, 2019.

\bibitem{thuy2013}
Pham~Thi Thuy, Young-Koo Lee, and Sungyoung Lee.
\newblock {A Semantic Approach for Transforming {XML} Data into {RDF}
  Ontology}.
\newblock {\em Wirel. Pers. Commun.}, 73(4):1387--1402, 12 2013.

\bibitem{thuy2009}
Pham Thi~Thu Thuy, Young-Koo Lee, and SungYoung Lee.
\newblock {{DTD2OWL}: Automatic Transforming XML Documents into {OWL}
  Ontology}.
\newblock In {\em {Proceedings of the 2Nd International Conference on
  Interaction Sciences: Information Technology, Culture and Human}}, {ICIS
  '09}, pages 125--131, New York, NY, USA, 2009. ACM.

\bibitem{tirmizi2008}
Syed~Hamid Tirmizi, Juan Sequeda, and Daniel Miranker.
\newblock {Translating SQL Applications to the Semantic Web}.
\newblock In {\em {Database and Expert Systems Applications}}, {DEXA '08},
  pages 450--464, Berlin, Heidelberg, 2008. Springer Berlin Heidelberg.

\bibitem{tomaszuk2010}
Dominik Tomaszuk.
\newblock {Flat triples approach to {RDF} graphs in {JSON}}.
\newblock In {\em {W3C Workshop -- RDF Next Steps}}. World Wide Web, 2010.

\bibitem{tomaszuk2011}
Dominik Tomaszuk.
\newblock {Named graphs in {RDF/JSON} serialization}.
\newblock {\em Zeszyty Naukowe Politechniki Gda{\'n}skiej}, pages 273--278,
  2011.

\bibitem{tomaszuk2012}
Dominik Tomaszuk, Karol Pak, and Henryk Rybinski.
\newblock {Trust in RDF Graphs}.
\newblock In {\em {ADBIS (2)}}, volume 186 of {\em {Advances in Intelligent
  Systems and Computing}}, pages 273--283. Springer, 2012.

\bibitem{tsinaraki2007}
Chrisa Tsinaraki and Stavros Christodoulakis.
\newblock {Interoperability of {XML} Schema Applications with {OWL} Domain
  Knowledge and Semantic Web Tools}.
\newblock In {\em {Proceedings of the 2007 OTM Confederated International
  Conference on On the Move to Meaningful Internet Systems: CoopIS, DOA,
  ODBASE, GADA, and IS - Volume Part I}}, {OTM'07}, pages 850--869, Berlin,
  Heidelberg, 2007. Springer-Verlag.

\bibitem{udrea2006}
Octavian Udrea, Diego~Reforgiato Recupero, and V.~S. Subrahmanian.
\newblock {Annotated RDF}.
\newblock In {\em {Proceedings of the 3rd European Conference on The Semantic
  Web: Research and Applications}}, {ESWC'06}, pages 487--501, Berlin,
  Heidelberg, 2006. Springer-Verlag.

\bibitem{Urbani2010}
Jacopo Urbani, Jason Maassen, and Henri Bal.
\newblock {Massive Semantic Web Data Compression with MapReduce}.
\newblock In {\em {Proceedings of the 19th ACM International Symposium on High
  Performance Distributed Computing}}, {HPDC '10}, pages 795--802, New York,
  NY, USA, 2010. ACM.

\bibitem{Urbani2013}
Jacopo Urbani, Jason Maassen, Niels Drost, Frank Seinstra, and Henri Bal.
\newblock {Scalable {RDF} data compression with MapReduce}.
\newblock {\em Concurrency and Computation: Practice and Experience},
  25(1):24--39, 2013.

\bibitem{vavliakis2013}
Konstantinos~N. Vavliakis, Theofanis~K. Grollios, and Pericles~A. Mitkas.
\newblock {{RDOTE} - Publishing Relational Databases into the Semantic Web}.
\newblock {\em J. Syst. Softw.}, 86(1):89--99, 1 2013.

\bibitem{verborgh2014}
Ruben Verborgh, Olaf Hartig, Ben {De Meester}, Gerald Haesendonck, Laurens {De
  Vocht}, Miel {Vander Sande}, Richard Cyganiak, Pieter Colpaert, Erik Mannens,
  and Rik {Van de Walle}.
\newblock {Querying Datasets on the Web with High Availability}.
\newblock In {\em {The Semantic Web -- ISWC 2014}}, pages 180--196, Cham, 2014.
  Springer International Publishing.

\bibitem{wu2006}
Zhaohui Wu, Huajun Chen, Heng Wang, Yimin Wang, Yuxin Mao, Jinmin Tang, and
  Cunyin Zhou.
\newblock {Dartgrid: a Semantic Web Toolkit for Integrating Heterogeneous
  Relational Databases}.
\newblock In {\em {Semantic Web Challenge at 4th International Semantic Web
  Conference}}, Athens, USA, 11 2006.

\bibitem{xiao2006}
Huiyong Xiao and Isabel~F Cruz.
\newblock {Integrating and Exchanging {XML} Data Using Ontologies}.
\newblock In {\em {Journal on Data Semantics VI}}, pages 67--89, Berlin,
  Heidelberg, 2006. Springer Berlin Heidelberg.

\bibitem{yahia2012}
Nora Yahia, Sahar~A. Mokhtar, and Amr Ahmed.
\newblock {Automatic Generation of {OWL} Ontology from {XML} Data Source}.
\newblock {\em CoRR}, abs/1206.0570, 2012.

\bibitem{yang2003}
Guizhen Yang and Michael Kifer.
\newblock {Reasoning about Anonymous Resources and Meta Statements on the
  Semantic Web}.
\newblock {\em J. Data Semantics}, 1:69--97, 2003.

\bibitem{Zimmermann2014}
Antoine Zimmermann.
\newblock {{RDF} 1.1: On Semantics of {RDF} Datasets}.
\newblock {W3C} note, World Wide Web Consortium, 2 2014.

\bibitem{straccia2010}
Antoine Zimmermann, Nuno Lopes, Axel Polleres, and Umberto Straccia.
\newblock {A General Framework for Representing, Reasoning and Querying with
  Annotated Semantic Web Data}.
\newblock {\em Journal of Web Semantics}, 11:72--95, 2012.

\end{thebibliography}

\end{document}